\documentclass{article}

\usepackage{arxiv}

\usepackage[utf8]{inputenc} 
\usepackage[T1]{fontenc}    
\usepackage{hyperref}       
\usepackage{url}            
\usepackage{booktabs}       
\usepackage{amsfonts}       
\usepackage{nicefrac}       
\usepackage{microtype}      
\usepackage{lipsum}
\usepackage{graphicx}
\usepackage[utf8]{inputenc}
\usepackage[english]{babel}
\usepackage[T1]{fontenc}
\usepackage{amsmath, amssymb,amsfonts, amsthm}
\usepackage{graphicx}
\usepackage{subcaption}
\usepackage{tabularx}
\usepackage{xcolor}

\newcommand{\argmin}{\mathrm{argmin}}

\newcommand\vv{\boldsymbol{v}}
\newcommand\R{\mathbb{R}}
\newcommand\qmes{q_{\mathrm{ref}}}
\newcommand\ymes{y_{\mathrm{ref}}}
\newcommand\vmes{v_{\mathrm{ref}}}
\newcommand\ysim{y_{\mathrm{sim}}}
\newcommand{\fig}[4]{\begin{figure}[h!]\centering\includegraphics[width=#1\linewidth]{./Figures/#2}\caption{#3}\label{#4}\end{figure}}
\usepackage[backend=biber]{biblatex}
\theoremstyle{definition}
\newtheorem{definition}{Definition}[section]
\newtheorem{assumption}{Assumption}[section]
\newtheorem{example}{Example}[section]

\title{Diffeomorphic registration distances for Bayesian calibration of infinite-dimensional computer models}

\author{
 Paul Lartaud \\
  CEA, DAM, DIF, F-91297 Arpajon, France \\
  \texttt{paul.lartaud@cea.fr} \\
   \And
 Gwenaël Salin \\
  CEA, DAM, DIF, F-91297 Arpajon, France
}
\begin{document}
\maketitle

\begin{abstract}
The simulation of physical phenomena with computer models relies on the estimation of physical and/or numerical parameters calibrated to fit experimental data. The approximations within the computer model and the errors in the measurements lead to uncertainties in the calibrated parameters. Bayesian calibration offers a well-studied framework to provide reliable uncertainty quantification on the calibrated parameters. \
When dealing with complex computer codes whose outputs are infinite-dimensional, Bayesian calibration may be extended by providing a relevant distance in the output space. In this paper, Bayesian calibration is performed using distances from the large deformation diffeomorphic metric matching (LDDMM) framework. LDDMM distances can provide a suitable metric for infinite-dimensional shapes such as scalar fields (\textit{i.e.} images) or function graphs. This metric can be interpreted as the minimal energy deformation required to transform one shape into another. As such, it provides a readily interpretable metric for Bayesian calibration. On top of this, the representation of the diffeomorphism group as an exponential transformation of an RKHS is compatible with Bayesian inference and allows to define a predictive posterior distribution on the infinite-dimensional space shape.
\end{abstract}

\section{Introduction}\label{sec:introduction}
This work deals with Bayesian calibration for computer models with infinite-dimensional outputs. Bayesian calibration describes a broad field of study whose goal is to estimate the parameters $\beta$ of a numerical model, with their uncertainties, to replicate experimental data of the physical phenomena to be modeled.

When the model outputs are infinite-dimensional (time series, functions, images, vector fields, …), the Bayesian calibration framework must be adapted, for example, with dimension reduction techniques \cite{higdon2008computer, conti2010bayesian, polette2025change}. It is also possible to use a kernel embedding and a Maximum Mean Discrepancy (MMD) metric in a pseudo-likelihood for the Bayesian inference problem \cite{cherief2020mmd}.

In this work, we propose a novel methodology based on the Large Deformation Diffeomorphic Metric Mapping (LDDMM) framework to provide interpretable distances in the infinite-dimensional output space.

In the LDDMM theory, one seeks to match one shape to another by deformation of the ambient space $E = \R^d$. The meaning of a shape will be further explained in Section \ref{sec:shape_space}, though for now the reader may understand it as a submanifold embedded in the ambient space. It is more general, however, as these shapes include, for example, curves and surfaces, images, or vector fields.

The ambient space deformations investigated are diffeomorphisms $\phi$ of the ambient space belonging to a subgroup $G_V$ of the group of diffeomorphisms $G = \mathrm{Diff}(E)$. This subgroup is defined as the set of flows generated from time-dependent velocity fields $\vv = (v_t)_{0 \leq t \leq 1}$ where $v_t \in V$ are smooth vector fields belonging to a Reproducing Kernel Hilbert Space (RKHS). The definition of the vector space $V$ and regularity conditions on the velocity fields are more thoroughly described in Section \ref{sec:distances}. The distance defined by the LDDMM framework is more physically relevant to compare shapes, as the regularity conditions enforced on the generating velocity fields lead to smooth diffeomorphic transformations. On the other hand, methods like optimal transport can lead to crossing of trajectories and do not necessarily come with spatial regularity conditions \cite{chodosh2015discontinuity}. Moreover, the particular RKHS structure of the Lie algebra of the diffeomorphism group in LDDMM is compatible with Gaussian process surrogate modeling, which is an essential step for the Bayesian inference of the parameters $\beta$.

\section{Context and previous work}\label{sec:context}
Analysis of curve registrations has been explored in recent years, for example, using elastic distances \cite{cheng2016bayesian, qian2014} and has been successfully applied to Bayesian calibration of computer models with functional outputs \cite{francom2025elastic}. In this paper, we propose to adapt Bayesian calibration to deformation metrics used in the context of large deformations within the LDDMM theory.

The LDDMM framework was first introduced in \cite{beg2005computing} and has been thoroughly investigated since. For a detailed description of the theoretical premises, the authors refer to \cite{younes2010shapes, miller2006geodesic, miller2002metrics}. This framework has also been formulated in optimal control theory \cite{arguillere2015shape, mang2017lagrangian}. This theory has been extended to the metamorphosis of images, in which variations of the shape are a mixture of deformation and source creation \cite{franccois2021metamorphic, holm2009euler, trouve2005metamorphoses}. This extension will not be discussed in this work.
A brief description of the LDDMM theory is provided hereafter.

\subsection{Distances between diffeomorphisms}\label{sec:distances}
Throughout this paper, we are working in the ambient space $E = \R^d$. The goal of LDDMM is to find deformations of the ambient space that match one source shape $q_0$ to a target shape $q_f$. The notion of \textit{shape} is deliberately left vague for now. For now, the reader may understand shapes as submanifolds of $\R^d$.

The deformations investigated belong to the set of diffeomorphisms of the ambient space denoted by $G = \mathrm{Diff}(E)$. More specifically, we seek diffeomorphisms as flows of ordinary differential equations, that is $\phi_{t=1}^{\vv}$ where $(\phi_t^{\vv})_{0 \leq t \leq 1}$ are solutions of the ordinary differential equation (ODE):
\begin{align}\label{eq:edo_flow}
\frac{d \phi_t^{\vv}}{dt} &= v_t \circ \phi_t^{\vv} \\
\phi_0^{\vv} &= \mathrm{Id}_{G}\nonumber
\end{align}
with $\vv = (v_t)_{0 \leq t \leq 1}$ a time-dependent velocity field. $\phi_t^{\vv}$ is the flow generated by the time-dependent velocity field $\vv$. Intuitively, the deformation $\phi_t^{\vv}$ acts on a point $x \in \R^d$ by transporting it along the velocity field.

Not all flows are diffeomorphisms of the ambient space. As an intuition, if the velocity field is not $C^1$, one may imagine a velocity field with a discontinuity at $y = 0$ such that $v(x, y) = (1, 1)$ for $y \geq 0$ and $v(x, y) = (1, -1)$ for $y < 0$. In this example, the flow $\phi_t$ at $t > 0$ is not injective: if $\phi_t(x, y) = (a, 0)$ then $x = a - t$ and $y = \pm t$.

Let $C_0^1(\R^d, \R^d)$ be the space of $C^1$ vector fields such that the vector field and its derivatives decrease to $0$ at infinity. This space is a Banach space for the norm $\| \cdot \|_{1, \infty}$ defined by:
\begin{equation*}
\| v \|_{1, \infty} = \mathrm{sup}_{x \in \R^d}\left\{|v(x)| + | \nabla v (x)| \right\}.
\end{equation*}

If $v_t \in C_0^1(\R^d, \R^d)$ for all $0 \leq t \leq 1$, the flow $\phi_1^{\vv}$ is a diffeomorphism. However, the Banach space $C_0^1(\R^d, \R^d)$ does not provide weak compactness, which is essential for the existence of minimizers in variational problems: bounded sequences of velocity fields may fail to have weak limits in $C_0^1(\R^d, \R^d)$.

To overcome this, velocity fields are chosen in an \textit{admissible} vector space $V \subset C_0^1(\R^d, \R^d)$. A vector space is called admissible if it is a Hilbert space with norm $\| \cdot \|_V$ continuously embedded in the space $C_0^1(\R^d, \R^d)$, that is, there exists a constant $C_V > 0$ such that for all $v \in V$, $\| v \|_{1, \infty} \leq C_V \| v \|_V$. The reason for this particular definition is made explicit in Section \ref{sec:matching_problem}.

Let $L_V^2 = L^2([0, 1], V)$. For $\vv = (v_t)_{0 \leq t \leq 1} \in L_V^2$, we introduce the norm $\| \vv \|_{L^2_V} = \int_0^1 \| v_t \|_V^2 dt$. $L_V^2$ equipped with the norm $\| \ \cdot \ \|_{L^2_V}$ is a Hilbert space of time-dependent velocity fields.

The admissible space $V$ is commonly chosen as an RKHS with kernel $K_V$. More generally, it can be defined by a scalar product $\langle u, v \rangle_V = \langle Lu, v \rangle_{L_2}$, with $L$ a pseudo-differential operator $L \colon C_0^1(\R^d, \R^d) \to C_0^1(\R^d, \R^d)$. The kernel $K_V$ is then the Green function associated with the operator $L$.

Let $G_V = \left\{\phi_1^{\vv} \ | \ \vv \in L_V^2 \right\} \subset \mathrm{Diff}(E)$. This subgroup is in fact equals to $\left\{\phi_t^{\vv} \ | \ \vv \in L^2_V \right\}$ since $\phi_t^{\vv} = \phi_1^{\mathbf{1}_{[0, t]} \vv }$ with $\mathbf{1}_A$ the characteristic function for the set $A$. In \cite{trouve1998diffeomorphisms}, $G_V$ is given the right-invariant weak Riemannian metric $\langle \cdot , \cdot \rangle_V$ and the associated Riemannian distance $d_G$ which is understood as the path length in the Riemannian manifold:
\begin{equation}\label{eq:dist_GV}
d_G(\mathrm{Id}_{G}, \phi) = \inf_{\vv \in L^2_V} \left\{ \| \vv  \|^2_{L_V^2} \ | \ \phi_1^{\vv} = \phi \right\}.
\end{equation}
It can be interpreted as the minimal energy deformation required to transform the ambient space. Here, the distance is defined relative to the identity, which leaves the ambient space unchanged. The right-invariance allows us to define a distance between any $\phi, \psi \in G_V$ by $d_G(\phi, \psi) = d_G(\mathrm{Id}_{G}, \psi \circ \phi^{-1})$.

$G_V$ can be identified as  Lie group whose tangent space at $\mathrm{Id}_{G}$ is $V$. $V$ is thus identified to the Lie algebra associated with the Lie group $G_V$. This identification is not rigorous as $V$ is not stable by the Lie bracket, which reduces the regularity of vector fields in $V$. Thus, $G_V$ is not a proper Lie group. Yet, the Lie group formalism is useful to derive the main results from LDDMM theory and it is used frequently throughout the literature.

Now, to compare two shapes, one can find $\phi \in G_V$ such that the deformed source shape matches the target shape and then record $d_G(\mathrm{Id}_{G}, \phi)$.

The natural question that arises then is: how do we compare shapes, and how does a deformation of the ambient space affect a shape?

\subsection{On shape representations}\label{sec:shape_space}
The LDDMM framework has been built to compare shapes, generally in 2D or 3D space. Most notably, it has been extensively used in computational anatomy \cite{trouve2005local, vialard2012diffeomorphic} to compare 2D or 3D images. A central question is thus how to best represent these shapes.

Formally, the representation of a shape requires three notions. First, we must define a shape space $\mathcal{S}$. Then we need to define the action of a diffeomorphism $\phi \in \mathrm{Diff}(E)$ on a representation $q \in \mathcal{S}$. This action is given in the form of a Fréchet differentiable (w.r.t. $\phi$) mapping $(\phi, q) \mapsto \phi \cdot q$. Finally, to compare two shape representations, we need a mapping $C \colon \mathcal{S} \times \mathcal{S} \to \R^+$ Fréchet differentiable w.r.t the first argument. This mapping is often known as the matching functional.

\begin{definition}
A shape space $\mathcal{S}$ is a Banach space that verifies the following conditions:
\begin{itemize}
\item There exists an action $\rho$ of the Lie group $G_V$ on $\mathcal{
S}$, that is $\rho$ is a map from $G_V \times \mathcal{S}$ to $\mathcal{S}$. This map is supposed to be Fréchet differentiable with respect to its first argument. For $\phi \in G_V$ and $q \in \mathcal{S}$, the group action is often denoted $\rho(\phi, q) := \phi \cdot q$.
\item It is equipped with a function $C \colon \mathcal{S} \times \mathcal{S} \to \R^+$, which is Fréchet differentiable with respect to its first argument.
\end{itemize}
A shape space is identified by the triplet $(\mathcal{S}, \rho, C)$. We may also use the same denomination to refer solely to $\mathcal{S}$.
Although the Fréchet differentiability is not technically necessary to prove the existence of a minimizing geodesic in the LDDMM variational problem, they are required to derive Euler-Lagrange and EPDiff equations, which is why we include these conditions in the definition of the shape space.

Given a shape $q$, we make the additional assumption that the orbit map $\phi \mapsto \phi \cdot q$ is a Riemannian submersion. This induces a weak Riemannian manifold structure on the orbit $G_V \cdot q$ and allow us to define the tangent and co-tangent spaces at $q$, denoted respectively by $T_q \mathcal{S}$ and $T_q^* \mathcal{S}$. 

Of course, this assumption may not be valid for all $q \in \mathcal{S}$, so in practice we make the implicit assumption that we are working in a subspace of the shape space that verifies this.
\end{definition}

\begin{example}\label{example:landmark}
The simplest example of shape representations is the \textit{landmarks} representation, in which a shape is simply represented by a finite number of points $q = (q_1, …, q_n)$, known as landmarks. The shape space is thus finite-dimensional and given by $\mathcal{S} = E^n$. The action of $\phi \in \mathrm{Diff}(E)$ on $\mathcal{S}$ is simply the action on each individual landmark and thus $\rho(\phi, q) = \left(\phi(q_1), …, \phi(q_n) \right)$. The data attachment term is generally a $L^2$ loss, such that for $x, y \in \mathcal{S}$:
\begin{equation*}
C(x, y) = \frac{1}{n} \sum\limits_{i=1}^n |x_i - y_i |_2^2.
\end{equation*}
\end{example}

\begin{example}\label{example:images}
A shape can also be a gray-level image, or in other words, a scalar field. The shape space is then $\mathcal{S} = C^1(E, \R)$. The action of $\phi \in \mathrm{Diff}(E)$ on an image $q \in \mathcal{S}$ is the pullback transport $\rho(\phi, q) = q \circ \phi^{-1}$. The standard data attachment term for images is also the $L^2$ loss:
\begin{equation*}
C(q_1, q_2) = \int_E (q_1(x) - q_2(x))^2 dx.
\end{equation*}
Colored images may be treated in a similar fashion in a shape space $\mathcal{S} = C^1(E, \R^3)$.
\end{example}

\begin{example}
Objects may also be represented as Radon measures $\mu \in \mathcal{S} = \mathcal{P}^2(E)$. The diffeomorphic action is then the pushforward of the measure $\rho(\phi, \mu) = \phi_{\#}\mu$. Data attachment for measures can be Wasserstein distances or Maximum Mean Discrepancies (MMD).
More generally, this framework can be extended to currents and varifolds, which include information on the tangent vectors of the shape. More details on current and varifold registration can be found in \cite{charon2013varifold}.
\end{example}

We only provide a few examples of the wide set of data structures that can be registered in the LDDMM framework. We could also mention vector fields, diffusion tensors, parametric curves and surfaces, and many others \cite{ceritoglu2009multi, charlier2017fshape, charon2014functional}.

\subsection{The diffeomorphic shape matching problem}\label{sec:matching_problem}
Consider a shape space $(\mathcal{S}, \rho, C)$. Matching the shape $q_0 \in \mathcal{S}$ to a target $q_f \in \mathcal{S}$ by a deformation $\phi_1^{\vv} \in G_V$ reduces to solving the following problem:
\begin{equation*}
\phi_1^{\vv}\in \argmin_{\psi_1^{\vv} \in G_V} C(\psi_1^{\vv} \cdot q_0, q_f).
\end{equation*}
This problem is inherently ill-posed in $\mathrm{Diff}(E)$ and minimizers may not be found (see \cite{younes2010shapes} for details). A way to circumvent this difficulty is to restrict the search to $G_V$ and regularize the problem with the distance $d_G$ on $G_V$ introduced in \eqref{eq:dist_GV}. Equivalently searching for the velocity fields $\vv$ instead of $\phi_1^{\vv}$, the regularized problem becomes:
\begin{align}\label{eq:opt_problem_with_constraints}
\vv \in \argmin_{L^2_V} \int_0^1 |v_t |^2_V dt + C(\phi_1^{\vv} \cdot q_0, q_f)  \
\text{subject to } \frac{d \phi_t^{\vv}}{dt} = v_t \circ \phi_t^{\vv}
\end{align}
If the functional $\vv \mapsto C(\phi_1^{\vv} \cdot q_0, q_f)$ is weakly continuous, there exists a minimizer for this regularized problem. This assumption is verified since we assumed the matching term was Fréchet differentiable w.r.t. its first argument.

Let $\mathcal{J}(\vv)$ be the regularized functional defined in \eqref{eq:opt_problem_with_constraints}. The sketch of the proof is the following: if $(\vv_n)_{n \in \mathbb{N}}$ is a sequence reaching the infimum of $\mathcal{J}$, it is bounded in $L^2_V$. By Banach-Alaoglu’s theorem, we may extract a weakly converging subsequence. We keep the same notation for this sequence and denote by $\vv$ its limit. Using the continuous embedding of $V$ in $C_0^1(\R^d, \R^d)$, one can then prove that $(\phi_1^{\vv_n})_{n \in \mathbb{N}}$ converges uniformly on compact sets to $\phi_1^{\vv}$. The continuity of the matching functional, and the weak lower semicontinuity of the Hilbert norm in $V$ allows to conclude that $\mathcal{J}(\vv_*) = \mathrm{inf}_{\vv \in L^2_V}\mathcal{J}(\vv)$.

The key conditions for the existence of minimizers are the continuous embedding $V \hookrightarrow C_0^1(\R^d, \R^d)$ and the Hilbert space structure of $V$.

In that case, the minimum is reached for a geodesic trajectory $(\phi_t^{\vv})_{0 \leq t \leq 1}$ in $G_V$, that is a trajectory minimizing the Riemannian path length (see for example \cite{younes2010shapes}). In what follows, we give insights into how this constrained problem reduces to a search for a geodesic in the space of diffeomorphisms $G_V$.

\subsection{Minimizing geodesics in $G_V$}\label{sec:geodesic_GV}

Let us first introduce some definitions.
\begin{definition}
Given a Lie group $G$, a Lagrangian is a function $\mathcal{L} \colon TG \to \R$ defined on the tangent bundle $TG$ of the Lie group $G$.

Let $(\phi, \dot{\phi}) \in TG$. If a Lagrangian is right-invariant w.r.t the action of the group $G$, one can write $\mathcal{L}(\phi, \dot{\phi}) = \mathcal{L}(\mathrm{Id}, \dot{\phi} \circ \phi^{-1})$. Since a velocity $v \in T_{\mathrm{Id}}G$ is precisely $v = \dot{\phi} \circ \phi^{-1}$ , the Lagrangian can be written using a \textit{reduced} Lagrangian $\ell \colon T_{\mathrm{Id}}G \to \R^+$ defined by $\ell(v) = \mathcal{L}(\mathrm{Id}, v)$.
\end{definition}
The reduced Lagrangian in the LDDMM framework is $\ell \colon V \to \R^+$ defined by $\ell(v) = \frac{1}{2} \langle v, v \rangle_V$.

\begin{definition}
Let $(\mathcal{S}, \rho, C)$ be a shape space. For a velocity field $v \in V$ in the Lie algebra of the Lie group $G_V$, the infinitesimal action of $v$ on a shape $q \in \mathcal{S}$ is defined as:
\begin{equation*}
v \cdot q := \frac{d}{dt}_{|t=0} \rho(\phi_{t}^{\vv}, q) \in T_q \mathcal{S}.
\end{equation*}
with $\vv = (v_t)_{0 \leq t \leq 1}$ is such that $v_0 = v$.
\end{definition}

\begin{definition}
For $q \in \mathcal{S}$, we define a canonical dual variable $\pi$ living in the cotangent space $\pi \in T_q^*\mathcal{S}$. This canonical variable is by definition a linear function acting on infinitesimal shape variations $\delta q \in T_q \mathcal{S}$.
\end{definition}

%
%
%
%

Given these definitions, we can tackle the problem at stake, which is the resolution of the constrained optimization problem \eqref{eq:opt_problem_with_constraints}.

Minimizing a function $f \colon \mathcal{X}\to\R$ defined on a Banach manifold $\mathcal{X}$ with a constraint given by a section $g$ of a Banach vector bundle $\mathcal{E} \to \mathcal{X}$ can be done by introducing a Lagrange multiplier as a section of the dual bundle $\mathcal{E}^*$ and finding saddle points of the unconstrained functional $h \colon  \mathcal{E}^* \to \R$ defined for $(x, \lambda) \in \mathcal{E}^*$ by:
\begin{equation*}
h(x, \lambda) = f(x) + \lambda(g(x)).
\end{equation*}

To ensure the existence of the Lagrange multipliers via the implicit function theorem, the following three assumptions are sufficient \cite{zeidler2013nonlinear}:
\begin{itemize}
\item There exists a minimizer $x \in \mathcal{X}$ of $f$ on the constraint set $g^{-1}(\{0\})$.
\item The constraint $g$ is Fréchet-differentiable at $x$ and its derivative is a surjective linear map onto the fiber $\mathcal{E}_{x}$. We denote the derivative by $Dg_{x} \colon T_{x}\mathcal{X} \to \mathcal{E}_{x}$.
\item Additionally, the kernel of $Dg_{x}$ is complemented in $T_{x} \mathcal{X}$, which ensures that the constraint manifold $g^{-1}(\{0\})$ is a smooth Banach submanifold.
\end{itemize}

In our case, the flow constraint is defined on the product space $\mathcal{X} = H^1([0, 1], G_V) \times L^2([0, 1], V)$. $G_V$ can be seen as a subset of an affine Banach space $G_V \subset \mathrm{Id}_G + C_0^1(\R^d, \R^d)$. The $H^1$-regularity of the paths $(\phi_t)_{0 \leq t \leq 1}$ thus stems from the $L^2$-regularity of $(v_t)_{0 \leq t \leq 1}$ and the continuous embedding $V \hookrightarrow C_0^1(\R^d, \R^d)$.

Since $g(\boldsymbol\phi,\boldsymbol v)_t = \dot\phi_t - v_t\circ\phi_t \in T_{\phi_t}G_V$, the constraint returns a section in the pullback bundle $\boldsymbol \phi^* TG_V$. Indeed, we recall that the tangent bundle $TG_V \to G_V$ is the set of couples $(\phi, v)$ such that $v \in T_\phi G_V$. Given the flow $\boldsymbol\phi \colon [0, 1] \to G_V$, the pullback bundle consists of pairs $(t, v)$ such that $v \in T_{\phi_t} G_V$. It maps the tangent bundle structure back on $[0, 1]$ using the flow trajectory $\boldsymbol\phi = (\phi_t)_{0 \leq t \leq 1}$.

The constraint thus takes values in the bundle $\mathcal{E} \to \mathcal{X}$ whose fiber at $(\boldsymbol \phi, \boldsymbol v)$ is the set of $L^2$ sections of the pullback bundle $\mathcal{E}_{(\boldsymbol \phi, \boldsymbol v)} = L^2([0, 1], \boldsymbol \phi^* TG_V)$.


The existence of a minimizer is a standard result from LDDMM theory. For the second assumption, consider variations $\delta \phi_t \in T_{\phi_t} G_V$ and $\delta v_t \in T_{v_t} V = V$. the differential of the flow constraint is the section $Dg_{(\boldsymbol\phi, \boldsymbol v)}[\boldsymbol \delta\boldsymbol \phi, \boldsymbol \delta\boldsymbol v]_t = \dot{\delta \phi_t} - (Dv_t \circ \phi_t) \delta \phi_t - \delta v_t \circ \phi_t \in T_{\phi_t}G_V$. Let $\boldsymbol{w} = (w_t)_{0 \leq t \leq 1}$, with $w_t \in T_{\phi_t}G_V$. Taking $\delta v_t = 0$ yields an ODE on $\delta \phi_t$:
\begin{equation}\label{eq:dg_ode}
\dot{\delta \phi_t} - (Dv_t \circ \phi_t) \delta \phi_t = w_t
\end{equation}
Regularity conditions on the velocity field $v_t$ are enough to justify the existence of the solution of the ODE, and hence the surjectivity of $Dg_{(\boldsymbol\phi, \boldsymbol v)}$ onto $\mathcal{E}_{(\boldsymbol\phi, \boldsymbol v)}$.

The kernel of $Dg_{(\boldsymbol\phi, \boldsymbol v)}$ is obtained from the solutions $(\boldsymbol\delta\boldsymbol\phi, \boldsymbol\delta  \boldsymbol v )$ of $Dg_{(\boldsymbol\phi, \boldsymbol v)}[\boldsymbol \delta\boldsymbol \phi, \boldsymbol \delta\boldsymbol v] = \boldsymbol 0$. Its complement can be constructed as follows: for any $(\boldsymbol \delta\boldsymbol\phi, \boldsymbol \delta\boldsymbol v )$, we compute first $\boldsymbol \delta \boldsymbol\phi^{(1)}$ as the unique solution of the ODE $Dg_{(\boldsymbol\phi, \boldsymbol v)}[\boldsymbol \delta\boldsymbol \phi^{(1)}, \boldsymbol \delta\boldsymbol v]_t= 0$ with initial condition $\delta \phi_0^{(1)} = \delta \phi_0$. Then we decompose $(\delta \phi_t, \delta v_t)$ as $(\delta \phi_t, \delta v_t) = (\delta \phi_t^{(1)}, \delta v_t) + (\delta \phi_t - \delta \phi_t^{(1)}, 0)$. By construction, $(\boldsymbol \delta \boldsymbol \phi^{(1)}, \boldsymbol \delta \boldsymbol v)  \in \mathrm{ker}(Dg_{\boldsymbol \phi, \boldsymbol v})$, and the second element $(\boldsymbol \delta \boldsymbol \phi - \boldsymbol \delta \boldsymbol \phi^{(1)}, \boldsymbol 0)$ defines the complement space. It is easy to check that their intersection is zero using the same argument as the uniqueness of the solution of \eqref{eq:dg_ode}.

The Lagrange multiplier is a section of the dual bundle whose fiber $\mathcal{E}^*_{(\boldsymbol \phi, \boldsymbol v)}$ at $(\boldsymbol \phi, \boldsymbol v)$ is a section of the dual pullback bundle $\mathcal{E}^*_{(\boldsymbol \phi, \boldsymbol v)} = L^2([0, 1], \boldsymbol \phi^* T^*G_V)$. It is a path in the cotangent spaces with $p_t \in T^*_{\phi_t}G_V$. The constrained optimization problem can then be seen as an unconstrained optimization problem for the augmented action $S$ defined by:
\begin{equation*}
S(\boldsymbol\phi, \boldsymbol p, \boldsymbol v) = \int_0^1 \left( \frac{1}{2} \langle v_t, v_t \rangle_V + \langle p_t, \dot{\phi_t}  - v_t \circ \phi_t \rangle \right) dt + C(\phi_1 \cdot q_0, q_f).
\end{equation*}
with $\langle \ \cdot \ , \ \cdot \ \rangle$ being the duality bracket. When necessary, we may add the relevant spaces as indices in the bracket notation.

To study this optimization problem, one may consider variations $\delta \phi_t \in T_{\phi_t}G_V$, $\delta p_t \in T_{p_t}T^*_{\phi_t}\mathcal{S} = T^*_{\phi_t}\mathcal{S}$ and $\delta v_t \in T_{v_t}V = V$. We denote by $\delta S_{\boldsymbol\phi}$, $\delta S_{p}$ and $\delta S_{\boldsymbol v}$ the action variations.

The optimality condition $\delta S_{\boldsymbol p} = \int_0^1 \langle \delta p_t, \dot{\phi_t} - v_t \circ \phi_t \rangle dt = 0$ for all $\delta p_t$ leads to the constraint \eqref{eq:edo_flow}.

Then, we can write with an integral by parts:
{\small
\begin{align*}
\delta S_{\boldsymbol\phi} &= \int_0^1 \langle p_t, \delta \dot{\phi_t} - (Dv_t \circ \phi_t) \delta \phi_t \rangle dt + \langle  D_{\phi}C(\phi_1 \cdot q_0, q_f), \delta\phi_1 \rangle \nonumber \\
&= \int_0^1 \left( \langle p_t, - (Dv_t \circ \phi_t) \delta \phi_t \rangle - \langle \dot{p_t}, \delta \phi_t \rangle \right)dt + \langle  D_{\phi}C(\phi_1 \cdot q_0, q_f), \delta \phi_1 \rangle + \langle p_1 , \delta \phi_1 \rangle = 0
\end{align*}
}
with $Dv_t$ the gradient of the vector field $v_t$ and $D_\phi C$ the Fréchet derivative of the cost functional with respect to its first argument.
This leads to the canonical equations:
\begin{equation}\label{eq:equation_pt}
\dot{p_t} + (D v_t(\phi_t))^* p_t = 0
\end{equation}
\begin{equation*}
p_1 = - D_{\phi}C(\phi_1 \cdot q_0, q_f).
\end{equation*}

Finally, let us write the variation $\delta S_{\boldsymbol v}$:
\begin{equation*}
\delta S_{\boldsymbol v} = \int_0^1 \langle Lv_t, \delta v_t \rangle - \langle p_t, \delta v_t \circ \phi_t \rangle dt = 0.
\end{equation*}
We recall that $L$ is the pseudo-differential operator defining the scalar product on the RKHS $V$.
The canonical variable is defined along the flow in $T_{\phi_t}G_V$ (in Lagrangian coordinates). We need to pull it back to Eulerian coordinates (in $T_{\mathrm{Id}_G} G_V$)  that are used to express the reduced Lagrangian $\ell(v_t) = \langle v_t, v_t \rangle_V$.
For $\phi \in G_V$, define $R\phi \colon \psi \mapsto \psi \circ \phi$. Then, we can notice that $\delta v_t \circ \phi_t$ is the pushforward of $\delta v_t$ by the right-translation operator $R_{\phi_t}$ at $\mathrm{Id}_G$, or equivalently $\delta v_t \circ \phi_t = D R{\phi_t}(\mathrm{Id}_G)(\delta v_t)$. Thus, we may simply use its dual action:
\begin{equation*}
\langle p_t, \delta v_t \circ \phi_t \rangle = \langle (D R{\phi_t}(\mathrm{Id}_G))^* p_t, \delta v_t \rangle.
\end{equation*}

Finally, the condition $\delta S_{\boldsymbol v} = 0$ leads to:
\begin{equation}\label{eq:pullback_pt}
L v_t = (D R_{\phi_t}(\mathrm{Id}_G))^* p_t.
\end{equation}
Introducing the \textit{Eulerian momentum} $m_t = (D R{\phi_t}(\mathrm{Id}_G))^* p_t \in V^*$, this last optimality condition can be written in a more concise form $L v_t = m_t$.

We may also write the dual action in $\R^d$-coordinates. Since $\delta v_t \circ \phi_t$ is a vector field, its dual element $p_t$ is a covector field (or a 1-form density). For $\phi \in G_V$, $v \in T_\phi G_V$ and $p \in T^*_\phi G_V$, and using the standard notation $\delta v(x) = \delta v^i(x) \partial_i$ and $p(x) = p_i(x) dx^i$ (with Einstein summation convention), the dual action is:
\begin{equation*}
\langle p, \delta v \circ \phi \rangle = \int_{E}  p_i(x) \delta v^i(\phi(x)) dx.
\end{equation*}
with $dx$ the usual Lebesgue measure on $\R^d$. A change of variable $y = \phi(x)$ leads to:
\begin{equation*}
\langle p, \delta v \circ \phi \rangle = \int_{E} |\mathrm{det} D\phi^{-1}(y) | p_i(\phi^{-1}(y)) \delta v^i(y) dy.
\end{equation*}
This expression gives us the explicit density $m_t(x)$ for $x \in \R^d$:
\begin{equation*}
m_t(x) = |\mathrm{det} D\phi^{-1}(x) | p_t(\phi^{-1}(x)).
\end{equation*}
The momentum $m_t$ is defined by the relation:
\begin{equation}\label{eq:euler_momentum}
\langle m_t, u \rangle = \langle p_t, u \circ \phi_t \rangle
\end{equation}
for all $u \in V$.

Let us introduce two more operators that are used to combine and simplify the optimality conditions.

\begin{definition}[Lie bracket on vector field]
The Lie bracket is a bilinear operator defined for $u, v \in V$ and for all $x \in E$ by:
\begin{equation*}
[u, v] = Dv(x) u(x) - Du(x) v(x)
\end{equation*}
where $Dv\colon E \to \R^d \times \R^d$ is the gradient of the velocity field $v$, and $Dv(x) u(x)$ is the matrix product. The Lie bracket $[u, v]$ may also be referred to as the \textit{adjoint operator} $\mathrm{ad}_u v$ or the \textit{Lie derivative} $\mathcal{L}_u v$.
\end{definition}

Intuitively, the Lie bracket measures the non-commutativity of the flows generated by the vector fields $u$ and $v$. The vector field $[u, v]$ can be understood as the velocity field of the infinitesimal variation of the vector field $v$ under transport by the flow generated by $u$.
\begin{definition}[Co-adjoint operator]
For a velocity field $v \in V$ the co-adjoint operator in $v$, denoted by $\mathrm{ad}_v^* \colon V^* \to V^*$ defined for $m \in V^*$ and $u \in V$ is defined by duality with the Lie bracket:
\begin{equation*}
\langle \mathrm{ad}_u^* m, v \rangle := \langle m, [u, v] \rangle
\end{equation*}
\end{definition}
The quantity $\mathrm{ad}_u^* m$ reflects how the dual quantity $m \in V^*$ is modified by an infinitesimal transport along the flow of $u$.

We now derive the EPDiff equation. Let us start by taking the time derivative of \eqref{eq:euler_momentum}:
\begin{align}
\langle \dot{m_t}, u \rangle &= \langle - (D v_t(\phi_t))^* p_t, u \circ \phi_t \rangle + \langle p_t, (Du \circ \phi_t) \dot{\phi_t} \rangle \nonumber \\
&= \langle p_t, (Du \circ \phi_t) v_t - (D v_t(\phi_t)) u \circ \phi_t \rangle \nonumber  \\
&= \langle m_t, (Du) v_t - (D v_t) u \rangle = \langle m_t, \mathrm{ad}_v u \rangle.   \nonumber
\end{align}
This last expression should be valid for all $u \in V$. Thus, using the co-adjoint operator $\mathrm{ad}^*_v$, we obtain the well-known EPDiff equation on the Eulerian momentum:
\begin{equation*}
\dot{m_t} + \mathrm{ad}_{v_t}^* m_t = 0.
\end{equation*}

Because the EPDiff equation on $m_t$ is deterministic, one only needs to know the initial momentum $m_0$ to derive the full trajectory. With $L v_t = m_t$, the flow equation \eqref{eq:edo_flow} and the canonical equation \eqref{eq:equation_pt}, the canonical trajectory $p_t$ and the diffeomorphism trajectory $\phi_t$ are also known. Solving the diffeomorphic registration problem is thus directly achieved by finding the initial canonical variable $p_0$. This is known as \textit{geodesic shooting}. Before introducing geodesic shooting, we will show that the previous derivation can be obtained equivalently in the shape space $\mathcal{S}$.

\subsection{Lagrangian formulation in shape space $\mathcal{S}$}\label{sec:geodesic_S}

Instead of writing the flow constraint $\dot{\phi_t} = v_t \circ \phi_t$ directly in $T_{\phi_t} G_V$, one may solve an optimization problem with a constraint in the shape space. Let $\boldsymbol q = (q_t)_{0 \leq t \leq 1}$ be the trajectory in the shape space. By definition of the infinitesimal action $v \cdot q$, the evolution of $q_t$ is governed by:
\begin{equation*}
\dot{q_t} - v_t \cdot q_t = 0.
\end{equation*}

Let us then introduce $\pi_t \in T_{q_t}^* \mathcal{S}$ the Lagrange multiplier associated to the shape constraint. The augmented action is then given by:
\begin{equation*}
S(\boldsymbol q, \boldsymbol\pi, \boldsymbol v) = \int_0^1 \left( \frac{1}{2} \langle v_t, v_t \rangle_V + \langle \pi_t, \dot{q_t}  - v_t \cdot q_t \rangle \right) dt + C(q_1, q_f).
\end{equation*}

We introduce the notation $\xi_{q_t}v_t = v_t \cdot q_t$. The dual of the infinitesimal action is  $\xi_{q_t}^*$. For $\pi_t \in T_{q_t}^* \mathcal{S}$, it is defined by its action $\langle\xi_{q_t}^*\pi_t, v_t \rangle_{V^*, V} =\langle \pi_t, \xi_{q_t}v_t\rangle_{T_{q_t}^* \mathcal{S}, T_{q_t} \mathcal{S}} = \langle \pi_t, v_t \cdot q_t\rangle_{T_{q_t}^* \mathcal{S}, T_{q_t} \mathcal{S}}$. We also introduce $D_{q_t}\xi_{q_t}v_t$ its Fréchet derivative taken at $q_t$. Following the reasoning of the previous section, we can derive three equations by considering variations $\delta q_t \in T_{q_t} \mathcal{S}$, $\delta \pi_t \in T_{q_t}^\mathcal{S}$ and $\delta v_t \in V$:
\begin{equation*}
\dot{q_t} - v_t \cdot q_t = 0
\end{equation*}
\begin{equation*}
\dot{\pi_t} + \left(D_{q_t}\xi_{q_t}v_t\right)^* \pi_t= 0
\end{equation*}
\begin{equation*}
L v_t - \xi_{q_t}^*\pi_t = 0.
\end{equation*}

This formulation allows us to work directly in the dual of the shape space. Although equivalent, this formulation may be easier to implement numerically. For instance, when working with images of size $N \times N$, geodesic shooting (see Section \ref{sec:hamilton_formalism}) in $(q, \pi)$-space amounts to an optimization problem on the $N \times N$ scalar field $\pi_0$, while the $(\phi, p)$ formalism leads to an optimization problem on $p_0$ of dimension $N \times N \times 2$.
It is not necessary to parametrize the full pixel grid. We can define a coarser grid of \textit{control points} and tune $p_0$ or $\pi_0$ defined on these control points \cite{durrleman2011optimal}.

\begin{example}
We consider the shape space of landmarks $\mathcal{S} = E^n$ defined in Example \ref{example:landmark}. For $v \in V$ and $q = (q_1, …, q_n) \in \mathcal{S}$, the infinitesimal action is $v \cdot q = (v(q_1), …, v(q_n)) \in E^n$. Its dual is defined by $\xi_q^*(\pi) = (v \cdot q)^T \pi$ for $\pi \in T_q^* \mathcal{S} = E^n$. The Fréchet derivative $D_{q}\xi_q v$ is given by the relation: $D_{q}\xi_q v (\delta q) = \left(Dv(q_1) \delta q_1, …, Dv(q_n) \delta q_n \right)$ for any $\delta q \in T_q \mathcal{S} = E^n$. The action of the dual is then $\left(D_{q}\xi_q v\right)^* \pi = \left(Dv(q_1)^T \pi_1, …, Dv(q_n)^T \pi_n\right)$.

Let $q_t, \pi_t \in T^\mathcal{S}$. The geodesic equations yield:
\begin{equation*}
\dot{q_t} + \mathbf{v}_t = 0
\end{equation*}
\begin{equation*}
\dot{\pi_t} - Dv_t(\mathbf{x})^T \pi_t = 0
\end{equation*}
with $\mathbf{v}_t = (v_t(x_1), …, v_t(x_n))$ and where we introduce the compact notation $Dv_t(\mathbf{x})^T \pi = (Dv_t(x_1)^T \pi_1, …, Dv_t(x_n)^T \pi_n)$.
\end{example}

\begin{example}
We consider the shape space of images defined in Example \ref{example:images}. For $v \in V$ and $q \in \mathcal{S}$, the infinitesimal action is $v \cdot q =- \nabla q \cdot v$ where $\nabla q \cdot v$ denotes the standard scalar product. Its dual is $\xi_q^*(\pi) = - \pi \nabla q$ for $\pi \in T_q^* \mathcal{S}$. The Fréchet derivative $D_{q}\xi_q v$ is given by the relation: $D_{q}\xi_q v (\delta q) = \nabla (\delta q) \cdot v$ for any $\delta q \in T_q \mathcal{S}$. Its dual is obtained by an integration by parts (with boundary terms vanishing):
\begin{align*}
\langle \pi, D_{q}\xi_q v(\delta q) \rangle &= \int_E \pi(x) \left( v(x) \cdot \nabla \delta q(x) \right) dx \\
&= \sum\limits_{i = 1}^d \int_E \pi(x) v_i(x) \frac{\partial \delta q}{\partial x_i}(x) dx \\
&= - \int_E \sum\limits_{i=1}^d \frac{\partial}{\partial x_i}(\pi v_i )(x) \delta q(x) dx \\
&= \langle \left(D_{q}\xi_q v\right)^*\pi, \delta q \rangle.
\end{align*}
By identification, we have $\left(D_{q}\xi_q v\right)^* \pi = - \mathrm{div}(\pi v)$.
Let $q_t, \pi_t \in T^*\mathcal{S}$. The geodesic equations yield:
\begin{equation*}
\dot{q_t} + \nabla q_t \cdot v_t= 0
\end{equation*}
\begin{equation*}
\dot{\pi_t} + \mathrm{div}(\pi_t v_t) = 0.
\end{equation*}
\end{example}

\subsection{Geodesic shooting with Hamilonian formalism}\label{sec:hamilton_formalism}
The equivalence between Lagrangian and Hamiltonian formulations is a deep and intricate subject. In this section, we provide insights into the main aspects of this dual theory. For a comprehensive review on the subject, readers may refer to \cite{miller2006geodesic}.

\begin{definition}
Let $\mathcal{L} \colon TG \to \R$ be a right-invariant Lagrangian on the Lie group $G$. The Hamiltonian $H \colon T^*G \to \R$ is defined as the Legendre transform of $\mathcal{L}$ expressed in the variables $(\phi, p)$, that is, in the cotangent bundle $T^*G$.
\end{definition}

The Hamiltonian of the unconstrained problem is obtained from the augmented Lagragian.

By definition, the Hamiltonian is the Legendre transform of $\mathcal{L}$:
\begin{align*}
H(\phi, p) = \langle p, \dot{\phi} \rangle - \mathcal{L}(\phi, \dot{\phi}) \text{ with } \dot{\phi} = v_t \circ \phi.
\end{align*}

By right-invariance of the Lagrangian, and pulling back $p$ to $T_{\mathrm{Id}} G_V$ as was done in \eqref{eq:pullback_pt}, the Hamiltonian can be written in a reduced form:
\begin{align*}
H(\phi, p) := h(m) = \frac{1}{2}\langle m, Km \rangle
\end{align*}
where we used the relation $L v = m$ or equivalently $v = Km$.

We can also express the Hamiltonian using the dual variable $\pi \in T_q^* \mathcal{S}$. A similar method leads to:
\begin{equation*}
H(q, \pi) = \frac{1}{2} \langle \xi_q^* \pi, K \xi_q^* \pi  \rangle.
\end{equation*}
This expression is in fact linked to the formulation in the $(\phi, p)$-space by the identification $m = \xi_q^* \pi$. These two formulations of the Hamiltonian give the geodesic equations either in $(\phi, p)$-space or $(q, \pi)$-space, in the form of Hamilton’s equations:
\begin{align}
\frac{d \phi_t}{dt} = \frac{\partial H}{\partial p_t} &\text{  ;  } \frac{d p_t}{dt} = -\frac{\partial H}{\partial \phi_t}\nonumber  \\
&\text{ or } \nonumber \\
\frac{d q_t}{dt} = \frac{\partial H}{\partial \pi_t} &\text{  ;  } \frac{d \pi_t}{dt} = -\frac{\partial H}{\partial q_t}.\nonumber
\end{align}

In order to solve the search for minimizing geodesics, the \textit{shooting} method is often used. It relies on the fact that the geodesic equations (either in Hamiltonian or Lagrangian formalism) are deterministic, given the initial conditions. Thus, if we have the initial conditions $(\phi_0, p_0)$ or  $(q_0, \pi_0)$, we can solve the equations and compute the action, with the matching term that requires $q_1$ (or $\phi_1$). Because the geodesic equations are deterministic, and since we know $q_0$ (or $\phi_0 = \mathrm{Id}_G$), the search for minimizing geodesics reduces to the search for the initial condition $p_0$ (or $\pi_0$).

The shooting method thus consists of finding the optimal initial momentum that minimizes the total action. This is usually done by automatic differentiation of the action with respect to the initial momentum. It can be done either in the $(\phi, p)$-space or the $(q, \pi)$-space.

After solving the minimization, the LDDMM distance is simply given by the Hamiltonian $H(q_0, \pi_0)$. This distance can then be used for Bayesian calibration of numerical simulations whose output space is the previously defined shape space $\mathcal{S}$.

This work aims to use the LDDMM distance within a Bayesian calibration framework. This distance provides an interpretable metric to compare a large class of infinite-dimensional objects. Before diving into our main contribution, we provide an overview of Bayesian calibration, whose goal is to identify simulation parameters in a computer model and provide an uncertainty quantification on said parameters. The estimated uncertainties are critical to guarantee the validity of complex phenomena based on multiple computer models.

\subsection{Bayesian calibration in finite-dimensional space}\label{sec:bayesian_calibration}
The task of Bayesian calibration refers to the statistical Bayesian inference of unknown model parameters. It is largely used when trying to fit the predictions of complex computer codes with experimental measurements. Let $\beta \in \Theta \subset \R^p$ be the model parameters. We assume these parameters to be real-valued, although Bayesian calibration has been extended to categorical parameters \cite{storlie2015calibration, congdon2005bayesian}.

Given some parameters $x \in \mathcal{X}$, we want to model a physical phenomenon $y \colon \mathcal{X} \to \R$. Experimental measurements are conducted for $n$ inputs $\mathbf{x} = (x_i)_{1 \leq i \leq n}$. They yield noisy measurements of $y(\mathbf{x}) = (y(x_i))_{1 \leq i \leq n}$ denoted by $\ymes(\mathbf{x}) = y(\mathbf{x}) + \varepsilon(\mathbf{x})$ where $\varepsilon(\mathbf{x})$ are additive measurement noises. Since the simulation does not necessarily reproduce the physical phenomenon accurately, we account for a model error $\eta$, parametrized by some $\xi$, such that $y(\mathbf{x}) = \ysim(\mathbf{x}, \beta) + \eta(\mathbf{x}, \xi)$.
This leads to:
\begin{equation}\label{eq:calibration}
\ymes(\mathbf{x}) = \ysim(\mathbf{x}, \beta) + \eta(\mathbf{x}, \xi) + \varepsilon(\mathbf{x}).
\end{equation}

The model error can be given a parametric form $\eta(\mathbf{x}, \xi) = \xi^T g(x)$ for some basis function $g$. It is also common to represent the unknown model error by an additive Gaussian process $\eta \sim \mathcal{GP}\left(m_\xi(x), k_\xi(x, x’)\right)$. This is the traditional approach described in \cite{kennedy2001bayesian} and largely investigated since (for example, in \cite{perrin2018calibration}). In both cases, we write $\xi$ for the parameters describing the model error. We seek to estimate jointly $\beta$ and $\xi$.

From a Bayesian perspective, a \textit{prior} distribution of density $p(\beta)$ is assigned to the unknown simulation parameters. It reflects expert knowledge or knowledge from other sources of information. Similarly, a prior distribution of density $p(\xi)$ is chosen for the model error parameters.

Given the relation \eqref{eq:calibration}, one can then define a likelihood $L(\ymes | \xi, \beta)$. The posterior distribution density $p(\beta, \xi | \ymes)$ can be obtained by Bayes’ theorem $p(\beta, \xi | \ymes) \propto p(\beta, \theta) L(\ymes | \xi, \beta) = p(\beta) p(\xi) L(\ymes | \xi, \beta) $. Although the density is only known up to a multiplicative constant, it is possible to sample the posterior distribution using Monte Carlo Markov Chain (MCMC) methods.

\section{LDDMM distances for Bayesian calibration}
In this section, we operate within a shape space $(\mathcal{S}, \rho, C)$ that serves as the output space of our numerical simulation. We further assume we have access to a dataset $\mathcal{D}_n = (\boldsymbol\beta, \boldsymbol q)$ with $\boldsymbol \beta = (\beta^{(i)})_{1 \leq i \leq n} \in \Theta^n$ some simulation parameters and $\boldsymbol q = (q^{(i)})_{1 \leq i \leq n} \in \mathcal{S}^n$ the corresponding output shapes.

We are given some experimental measurement in the form of a shape $\qmes \in \mathcal{S}$. For the sake of concision, we assume a unique measurement, although the Bayesian framework can be easily adapted to account for multiple measurements. Our goal is to calibrate the simulation parameters and estimate the uncertainties on those parameters. Thus, we seek to obtain the posterior distribution $p(\beta | \mathcal{D}_n, \qmes)$.

\subsection{Bayesian formalism in the Lie algebra}\label{sec:prob_model}
In a Bayesian setting, the model parameters $\beta$ are random variables with a given prior density $p(\beta)$. In our applications, the output only depends on the parameters of interest $\beta$. For this reason, we remove the variable $x$ from our notation for concision. For $\beta \in \Theta$, let $f(\beta)$ be the output shape. Throughout this section, we make the following assumption:
\begin{assumption}\label{assum:in_orbit}
For all $\beta \in \Theta$, the shape $f(\beta)$ belongs to the orbit of $\qmes$ under the action of $G_V$.
\end{assumption}
Given this assumption, to each shape $f(\beta)$ corresponds an initial velocity field $v_0(\beta)$ yielding the registration of $f(\beta)$ to $\qmes$.

To calibrate the model, we want $v_0(\beta)$ to be close to $0$: if $v_0 = 0$, the geodesic in $G_V$ is $\phi_t^{\vv} = \mathrm{Id}_G$ for all $0 \leq t \leq 1$.

We account for an additive noise term with an infinite-dimensional probabilistic model. Let $\varepsilon$ be the Gaussian measure with zero-mean and covariance $C_V$ being the integral operator associated to $K_V$:
\begin{equation*}
C_V f = \int_{\R^d} K_V( \ \cdot \, y) f(y) dy.
\end{equation*}

The RKHS $V$ is the Cameron-Martin space \cite{gross1967abstract, kuo2006gaussian} of the Gaussian measure $\varepsilon$. This measure is supported in a larger space given by the closure of $V$.

Let us define the following probabilistic model:
\begin{equation*}
\vmes = v_0(\beta) + \varepsilon
\end{equation*}
where $\vmes$ is initial the velocity field transporting $\qmes$ to $\qmes$, which is by construction $\vmes = 0$. 

In what follows, if $X$ is a random variable, we denote by $\mathbb{P}_X$ its probability measure.

Since the velocity field $v_0(\beta)$ lives in the Cameron-Martin space $V$, by Cameron-Martin theorem, $\vmes$ is absolutely continous with respect to the Gaussian measure $\varepsilon$ and the Radon-Nykodim derivative is known. Furthermore, Bayes' theorem holds in infinite dimension in the sense that the measure of $\beta | \vmes$ is absolutely continuous with respect to the prior measure on $\beta$ and the Radon-Nykodim derivative is given by:
\begin{equation*}
\frac{d\mathbb{P}_{\beta | \vmes}}{d\mathbb{P}_{\beta}} = \frac{1}{Z}\exp\left(-\frac{1}{2} \|\vmes - v_0(\beta) \|_V^2 \right).
\end{equation*}
In our case, by construction $\vmes = 0$. Thus we can define an infinite-dimensional analog to the likelihood by:

\begin{equation*}
L(\qmes | \beta) := \frac{d\mathbb{P}_{\beta | \vmes=0}}{d\mathbb{P}_{\beta}} = \frac{1}{Z}\exp\left(-\frac{1}{2} \|v_0(\beta) \|_V^2 \right)
\end{equation*}
with $Z$ a normalization constant. More details on this construction can be found in \cite{stuart2010inverse}.

%
%

%


The reason for the error term $\varepsilon$ is two-fold: first of all, there is no guarantee that there exists $\hat{\beta}$ such that $v_0(\hat{\beta}) = 0$. This is especially true if the model has a systematic bias compared to the experiment and cannot capture the whole shape structure. On top of this, the optimization problem in the geodesic shooting may not reach the global minimum, which would lead to some small errors in the estimation of $v_0$.

Given a measurement $\qmes$, our probabilistic model is summarized by the following log-likelihood:
\begin{equation}\label{eq:log_like_calib}
\log  L(\qmes | \beta) = - \frac{1}{2} \| v_0(\beta) \|^2_V + \mathrm{cst}
\end{equation}
where we put the normalization constant terms in $\mathrm{cst}$.

This likelihood defines a Bayesian probabilistic model for $\beta$. From a given measurement $\qmes$, we can obtain the posterior distribution $p(\beta | \qmes) $. This is done numerically with MCMC methods.

The posterior distribution $p(\beta | \qmes)$ can be used to define a posterior predictive distribution in the shape space. Consider the mapping $\psi \colon \beta \mapsto \phi_1^{\vv(\beta)} \cdot \qmes$. This mapping is measurable since $\beta \mapsto \vv(\beta)$ and $\vv \mapsto \phi_t^{\vv}$ are continuous.
A posterior predictive distribution can then be defined on $G_V \cdot \qmes$ by the pushforward of $\psi$ on the posterior distribution $p( \ \cdot \ | \qmes)$. This distribution lives in an infinite-dimensional space and typically does not have a density.

\subsection{Surrogate modeling in the tangent space}\label{sec:prob_model_sm}
To obtain this posterior distribution, we need to have access to the mapping $\beta \mapsto v_0(\beta)$. This mapping is not directly accessible, but we can approximate it with a surrogate model.

The first step is to evaluate the distances $d^{(i)} = d_G(q^{(i)}, \qmes)$ between each output of the simulation and the reference measurement. We also denote by $v_0^{(i)}$ the corresponding initial velocity fields. This step is done using the geodesic shooting approach described in Section \ref{sec:hamilton_formalism}.

We thus obtain a new database $(\boldsymbol\beta, \boldsymbol v_0)$ with $\boldsymbol v_0= (v_0^{(i)})_{1 \leq i\leq n}$ where $v_0^{(i)} \in V$.

To obtain the link between the simulation parameters and the deformations, we approximate the initial velocity field $v_0$ by $\widehat{v_0}$ defined as a $V$-valued zero-mean Gaussian process with an operator-valued kernel $K:\Theta \times \Theta \to B(V)$. We construct $K$ by tensorization with the covariance operator $C_V$.

Let $k: \Theta \times \Theta \to \R$ be a scalar-valued positive-definite kernel (Gaussian, Matérn, Sobolev ...). Then we define $K$ by:
\begin{equation*}
K(\beta, \beta') = k(\beta, \beta') C_V \text{ for } \beta, \beta' \in \Theta.
\end{equation*}

This operator-valued kernel $K$ is positive definite and self-adjoint, and for any $\beta, \beta' \in \Theta$ it returns a trace-class operator of $V$ (since $C_V$ is trace-class) \cite{kadri2016operator}. It defines a $V$-valued Gaussian process $\widehat{v_0} \sim \mathcal{GP}\left(0, K\right)$. In particular, it is possible to condition $\widehat{v_0}$ by the observations $\mathcal{D}_n = (\boldsymbol\beta, \boldsymbol v_0)$. The conditional process is also a Gaussian process, whose mean and kernel are  given by extension of the kriging equations.
\begin{gather*}
\widehat{v_0} | \mathcal{D}_n \sim \mathcal{GP}\left(v_\mathrm{mod}, K_c \right) \\
v_\mathrm{mod}(\beta) = K(\beta, \boldsymbol\beta) K(\boldsymbol\beta, \boldsymbol\beta)^{-1} \boldsymbol v_0 \\
K_c(\beta, \beta') = k_c(\beta, \beta') C_v = \left(k(\beta, \beta') -  K(\beta, \boldsymbol\beta) K(\boldsymbol\beta, \boldsymbol\beta)^{-1} K(\boldsymbol\beta, \beta') \right) C_V
\end{gather*}
where $K(\boldsymbol\beta, \boldsymbol\beta) = (K(\beta^{(i)}, \beta^{(j)}))_{1 \leq i, j \leq n}$, $K(\beta, \boldsymbol\beta) = (K(\beta, \beta)^{(i)})_{1 \leq i \leq n}$ and $K(\boldsymbol\beta, \beta) = K(\beta, \boldsymbol\beta)^T$.

Conditioning $\widehat{v_0}$ by the dataset $\mathcal{D}_n = (\boldsymbol\beta, \boldsymbol v_0)$ leads to a posterior Gaussian process $\widehat{v_0} | \mathcal{D}_n$. The mean function $\mathbb{E}[\widehat{v_0} | \mathcal{D}_n]$ should provide a good approximation for $v_0(\cdot)$. From now on, we introduce the notation $v_{\mathrm{mod}} = \mathbb{E}[\widehat{v_0} | \mathcal{D}_n]$.

We modify our initial probabilistic model to:
\begin{equation*}
\vmes = (\widehat{v_0} | \mathcal{D}_n)(\beta) + \varepsilon = v_{\mathrm{mod}}(\beta) + \varepsilon_{\mathrm{mod}}(\beta) + \varepsilon
\end{equation*}
where $\varepsilon_{\mathrm{mod}}$ is a zero-mean Gaussian process with kernel $K_c$ and $\varepsilon_{\mathrm{mod}}(\beta)$ is independent of $\varepsilon$ for $\beta \in \Theta$. We have by construction $v_{\mathrm{mod}}(\beta) \in V$ and $\varepsilon_{\mathrm{mod}}(\beta)$ a Gaussian measure on $V$. The independent sum $\varepsilon + \varepsilon_{\mathrm{mod}}(\beta)$ yields a zero-mean Gaussian measure with covariance operator $(1 + k_c(\beta, \beta)) C_V$.

By Cameron-Martin theorem, we can thus define the Radon-Nykodim derivative for this new model:
\begin{equation*}
\frac{d\mathbb{P}_{\beta | \vmes=0}}{d\mathbb{P}_{\beta}} = \frac{1}{Z'}\exp\left(-\frac{\|v_{\mathrm{mod}}(\beta) \|_V^2}{2(1 + k_c(\beta, \beta))} \right).
\end{equation*}

Introducing the GP uncertainty smoothes out the likelihood in regions of high uncertainties. Reciprocally, if the uncertainty of the surrogate is small (for example if we are close to a datapoint $\beta^{(i)}$), the likelihood tends to increase.

\subsection{Systematic error model}\label{sec:prob_model_disc}

Up until now, we assumed $\varepsilon$ was a centered Gaussian variable. This is not necessarily the case. In particular, there may not be any $\beta \in \Theta$ such that $f(\beta)$ approaches $\qmes$.

To account for a possible failure of the assumption, we may refine the surrogate model by adding a parametric mean function $m \colon \Theta \to V$ acting as a bias. Since the surrogate models use a PCA dimension reduction step, a simple discrepancy would be to consider an additive term linear with respect to the latent component of $u_{\mathrm{mod}}(\beta) = \mathrm{PCA}(v_{\mathrm{mod}}(\beta))$. The notation $ \mathrm{PCA}(v_{\mathrm{mod}}(\beta))$ refers to the PCA decomposition of the velocity field $v_{\mathrm{mod}}(\beta)$. The latent space dimension is denoted by $P$.

There is no particular reason that the additive discrepancy in the latent space should be linear in $\beta$, however, and this method has not proven to be relevant in our testing.

Using the dataset registrations, we know which simulation $q^{(i)}$ is the closest (in the sense of the LDDMM distance) to the measure $\qmes$. Its corresponding initial velocity field is denoted by $v_{\mathrm{min}}$ and the latent space representation is $u_{\mathrm{min}} =  \mathrm{PCA}(v_{\mathrm{min}})$.

For $(\xi_j)_{1 \leq j \leq P}$, we propose a discrepancy of the form:

\begin{equation}\label{eq:discrepancy}
u(\beta, \xi) = u_{\mathrm{mod}}(\beta) - \xi  u_{\mathrm{min}}.
\end{equation}

The reasoning is the following: the available data hints that the minimal deformation reachable, from $\qmes$ to a simulation $q^{(i)}$, is close to the deformation generated by the geodesic of initial velocity field $v_{\mathrm{min}}$.

By introducing \eqref{eq:discrepancy}, we should be able to reach deformation close to the identity for $\xi$ close to $1$, which should allow us to model the discrepancy of the simulation. We add a linear dependency in $\xi$ to allow for Bayesian inference with the discrepancy.

As in standard Bayesian calibration with additive discrepancies, however, we may lose some identifiability since Bayesian inference on $\xi$ may overlap with uncertainties on $\beta$. To reduce the effect of this phenomenon, we take narrow priors for $\xi = (\xi_j)_{1 \leq j \leq P}$, centered around one to limit those compensation effects.

Despite the partial loss of identifiability, the linear discrepancy may be useful to quantify (with uncertainties) the bias between experiment and simulation in the shape space. This will be highlighted in an example in Section \ref{sec:application_functions}.

\section{Applications}
In this section, we describe two examples of application. The first one is a toy case to highlight the feasibility of Bayesian calibration with image outputs. The second is an applied problem for the Bayesian calibration of Johnson's damage model with functional outputs.

For practical implementation, this work focuses on a Hamiltonian formalism of the geodesic equations in the $(q, \pi)$-space, implemented in PyTorch, and the minimization of the action functional (\textit{i.e.} the loss) is done by automatic differentiation. The forward numerical integration of Hamilton equations is done with an Euler-Maruyama scheme (otherwise known as \textit{leapfrog} integration) \cite{duane1987hybrid, neal1992bayesian}, designed to preserve the symplectic structure of the equations.

Additionally, after registering the measurement $\qmes$ to each simulation $q^{(i)}$, we also filter out a fraction of the simulations. Our criterion is to remove the $10 \%$ of the simulations with the largest LDDMM distances between $q^{(i)}$ and $\qmes$. This left-out data would not have been relevant for calibration, and removing it entirely facilitates the training of the GP-PCA surrogate model as it reduces the complexity of the high-dimensional outputs, which means fewer PCA components are required to explain a similar fraction of total variance. Regarding the PCA, we choose to keep enough PCA components to explain $95 \%$ of the data variance.

All the experiments were conducted on a single CPU 13th Gen Intel\textregistered Core\texttrademark i7-13850HX × 28.

\subsection{Toy case for Bayesian calibration of image outputs}
We start with a simple case. Consider a simulation code $f(\beta)$ that returns a $32 \times 32$ image where the (non-normalized) pixel intensity is given by
$p[i, j] = \exp\left(- 6 ((x_{ij} + \beta_1 \mathrm{sin}(2\pi\beta_3 y_{ij})^2 + (y_{ij} + \beta_2 \mathrm{cos}(2 \pi\beta_4 x_{ij})^2 )  \right)$ where $x_{ij}$ and $y_{ij}$ refer to the $x$ and $y$ coordinates of the pixel $[i, j]$.

Consider a measurement $\qmes$, which for the sake of simplicity, is simply one output of the code for $\beta_* = (0.2, 0.3, 0.4, 0.8)$. We seek the posterior distribution of $\beta$ given $\qmes$, using the LDDMM distance for calibration.

We use a dataset consisting of $n = 300$ samples $(\boldsymbol\beta, \boldsymbol q)$. The samples $\boldsymbol\beta$ are uniformly distribution within the bounds $[0, 0, 0, 0]$ and $[0.5, 0.5, 0.7, 0.7]$ and are sampled with Latin Hypercube Sampling.

The reference measurement $\qmes$ and examples of simulation images are highlighted in Figure \ref{fig:example_images}. We also present the mean and standard deviation of the images obtained from the prior on $\beta$ in Figure \ref{fig:prior_summary}.

\fig{0.99}{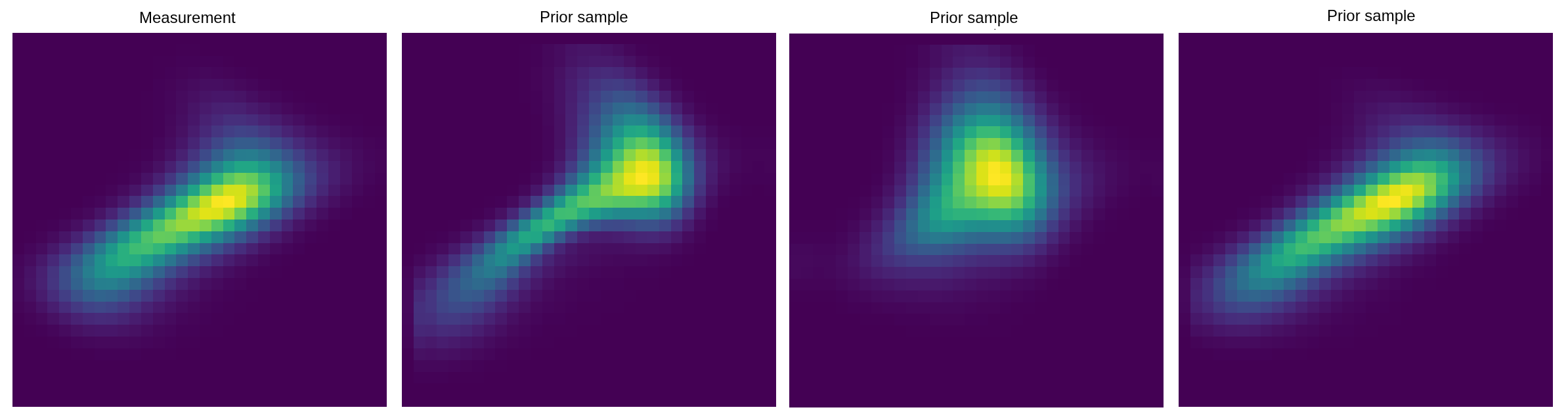}{Reference measurement (leftmost) and example of prior images.}{fig:example_images}

\fig{0.99}{prior_summary.png}{Mean (center) and standard deviation (right) of the prior predictive distribution in image space, compared to the reference measurement (left). Images are scaled between $0$ and $1$.}{fig:prior_summary}

For each simulation $q^{(i)}$, we solve the diffeomorphism matching problem by geodesic shooting, to transport $\qmes$ to $q^{(i)}$. $V$ is the RKHS associated with a Gaussian kernel with lengthscale $\sigma_V = 2 / 32$. The geodesic equations are solved with a second-order Runge-Kutta numerical scheme. The initial momentum is optimized by Adam \cite{kingma2014adam} using automatic differentiation in PyTorch, with gradient clipping to improve stability.

In Figure \ref{fig:image_registration}, we display an example of the geodesic found for the diffeomorphic registration of one simulation.

\fig{0.9}{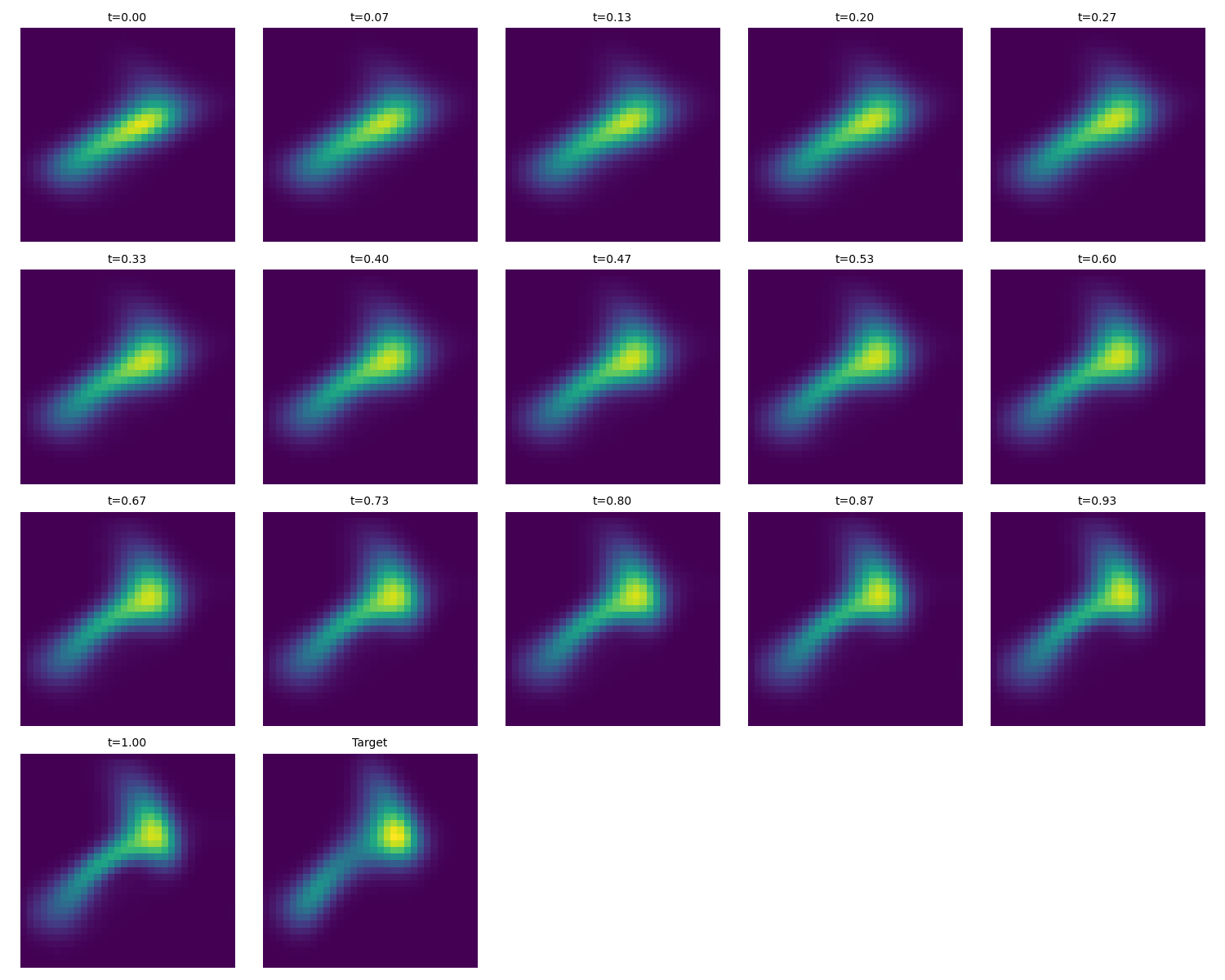}{Geodesic trajectory between a source image (upper left) and target image (lower right).}{fig:image_registration}

After a filtration of the $10 \%$ worst simulations, the initial momenta $\boldsymbol\pi_0$ are recorded for each simulation, and they are used to learn a surrogate model of the mapping $\beta \mapsto \pi_0(\beta)$ using PCA-based Gaussian Process Regression.

Some performance metrics for the GP regression are presented in Table \ref{tab:gp_pca_image}.


\begin{table}[h]
\centering
\begin{tabular}{lc}
\hline
Validation metric \qquad &  \\
\hline
Aggregated NRMSE \qquad & \qquad $0.187$ \qquad \\
Aggregated NMAE  \qquad & \qquad $0.132$ \qquad \\
Aggregated $Q^2$ \qquad & \qquad $0.936$ \qquad \\
Aggregated CRPS  \qquad & \qquad $0.042$ \qquad \\
Aggregated IAE   \qquad & \qquad $0.081$ \qquad \\
RRMSE (\%)       \qquad & \qquad $14.48$ \qquad \\
\hline
\end{tabular}
\caption{Predictive performance of the GP-PCA surrogate model for the image toy case.}
\label{tab:gp_pca_image}
\end{table}

The aggregated metrics are standard metrics averaged over the output dimensions. The Integrated Absolute Error (IAE) (see \cite{marrel2024probabilistic}) is evaluated in the true space after PCA reconstruction. We recall its definition:
\begin{equation*}
\mathrm{IAE}_d = \int_0^1 |\mathcal{P}_d(\alpha) - \alpha | d\alpha
\end{equation*}
with $\mathcal{P}_d(\alpha)$ the fraction of test points within the credible interval of level $\alpha$ for output dimension $d$. We then averaged the $\mathrm{IAE}_d$ over the output dimension.

Using the log-likelihood from \eqref{eq:log_like_calib}, we can perform Bayesian calibration using the LDDMM distances between images. We run a Hamiltonian Monte Carlo (HMC) sampling \cite{girolami2011riemann} with No U-Turn Sampler (NUTS) \cite{hoffman2014no}, using Pyro \cite{bingham2018pyro}. The marginal posterior distributions are highlighted in Figure \ref{fig:posterior_marginals}. The posterior distribution in $\beta$ defines a posterior predictive distribution on images. We display the MAP and the relative standard deviation of the predictive distribution in Figure \ref{fig:map_image}.

\fig{0.99}{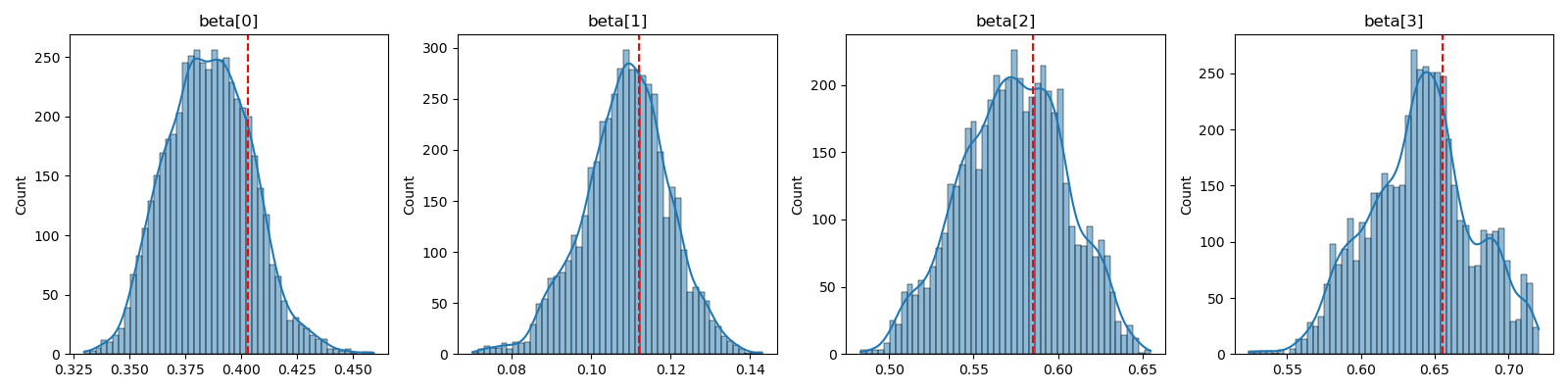}{Marginal posterior distributions of $p(\beta | \qmes)$. The red vertical line indicates the true value of $\beta_*$ used for the reference measurement.}{fig:posterior_marginals}

\fig{0.99}{posterior_summary.png}{Mean (center) and standard deviation (right) of the posterior predictive distribution in image space, compared to the reference measurement (left). Images are scaled between $0$ and $1$.}{fig:posterior_summary}

\fig{0.9}{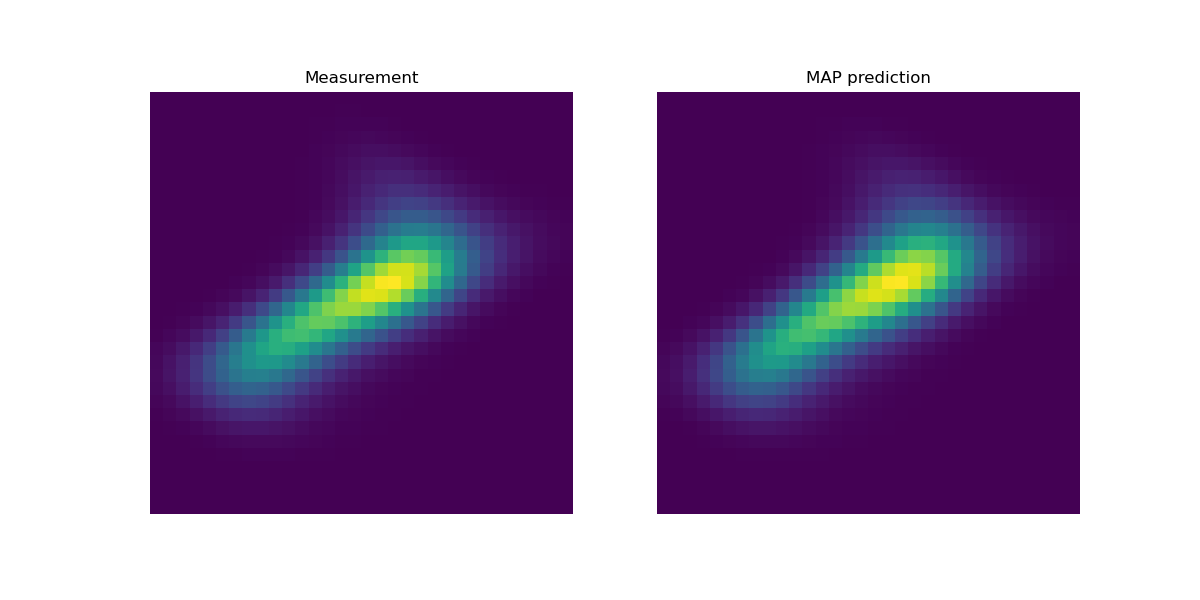}{Maximum a posteriori obtained during MCMC sampling of the posterior $p(\beta | \qmes)$.}{fig:map_image}

The predictive posterior distributions obtained display significantly smaller uncertainties. The posterior distribution $p(\beta | \qmes)$ encompasses the true value $\beta_*$. The MAP is not perfectly aligned with $\beta_*$. This may be linked to the rather small influence of $\beta_1$ in this example. This is illustrated by the strong similarities between the reference $\qmes$ and the MAP prediction in Figure \ref{fig:map_image}. Despite a bias in the parameters, the MAP prediction is very close to the reference.

\subsection{Functional Bayesian calibration of Johnson’s damage model}\label{sec:application_functions}
In this second application, we consider a simulation model $f \colon \mathcal{X} \times \Theta \to \mathcal{S}$ where $\Theta \subset \R^p$ and $\mathcal{S} = C^1([0, 1], [0, 1])$. Given a set of parameters $\beta \in \Theta$ and some inputs $x \in \mathcal{X}$, the simulation code $f(x, \beta)$ returns the time-evolution of the free surface velocity of a metallic plate impacted by a known projectile. This example is used to investigate the material properties of the plate under hydrodynamic conditions. Namely, we seek to identify parameters of Johnson’s model for plastic deformation.

We are given an experimental measurement of the plate response $q^{(\mathrm{mes})}$ and a set of simulations and their associated input parameters $\mathcal{D}_n = (\beta^{(i)}, q^{(i)})_{1 \leq i \leq n}$. The simulation curves and the experimental measurements are represented in Figure \ref{fig:simulation_curves}. Thus, our objective is to adapt Bayesian calibration to recover a posterior distribution $p(\beta | \mathcal{D}_n)$ using the LDDMM distance.

\fig{0.7}{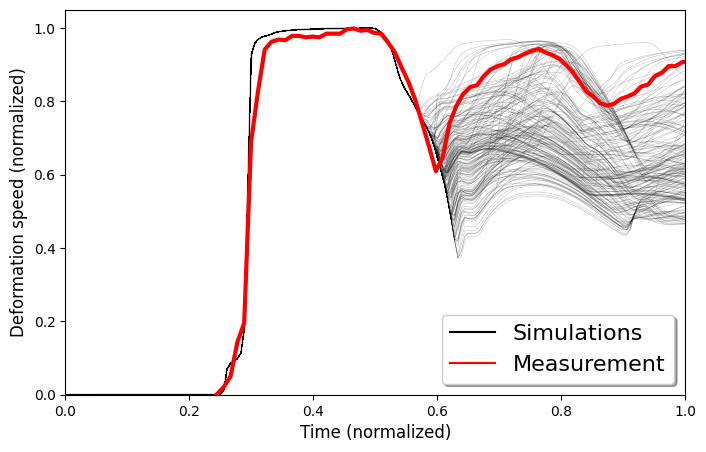}{Experimental measurement and simulation curves}{fig:simulation_curves}

The time resolution is different for the simulations and experimental measurements. Simulation curves are discretized over $n_{t_1} = 182$ timesteps, while experimental measurements are discretized over $n_{t_2} = 70$ timesteps.

The functional outputs are represented as currents here to include information on the tangential space of the curve. The matchings for the measure representation were very close.

To begin with, we performed standard Bayesian calibration with L2 loss using a GP-PCA surrogate model to learn the mapping between $\beta$ and the simulation curves. The MAP and $95 \%$ credible intervals of the predictive posterior distribution are displayed in Figure \ref{fig:predposterior_L2}.


\fig{0.7}{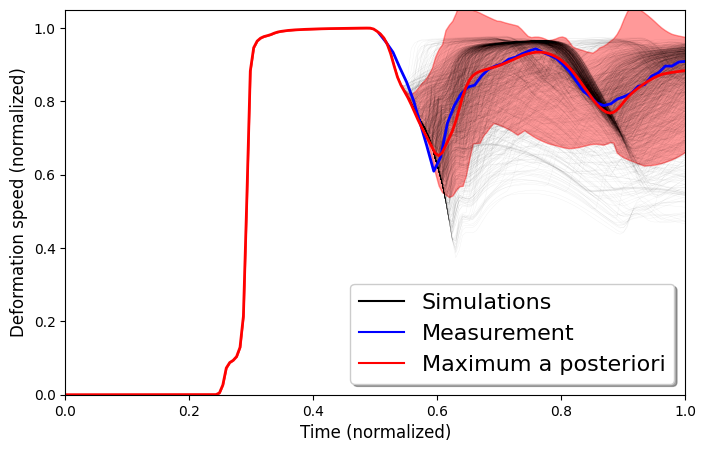}{Predictive posterior distribution and maximum a posteriori for the functional calibration of deformation speed with L2 distance and GP-PCA surrogate model.}{fig:predposterior_L2}

We see that we have very large residual uncertainties. The main cause is likely the use of the L2 loss, which is not readily suited to compare functional outputs with structural differences such as location and amplitude of extremal positions. The simulations and the reference measurement have some systematic differences, mainly in the time and amplitude of the first local minimum. We expect this bias to explain the majority of the L2 loss between a simulation and the reference measurement, meaning the simulation parameters are mostly non-informative in this L2 calibration problem. Consequently, the posterior distribution of $\beta$ is not narrowed down, and the posterior predictive uncertainties remain very large.

In what follows, we use our novel approach based on the diffeomorphic registration distance.

An example of curve matching for one simulation is presented in Figure \ref{fig:plot_appariement_curves}. We also present the deformation of ambient space and the Hamiltonian value, which is exactly the deformation energy $\frac{1}{2} \|v_0\|^2_V$.

\fig{0.99}{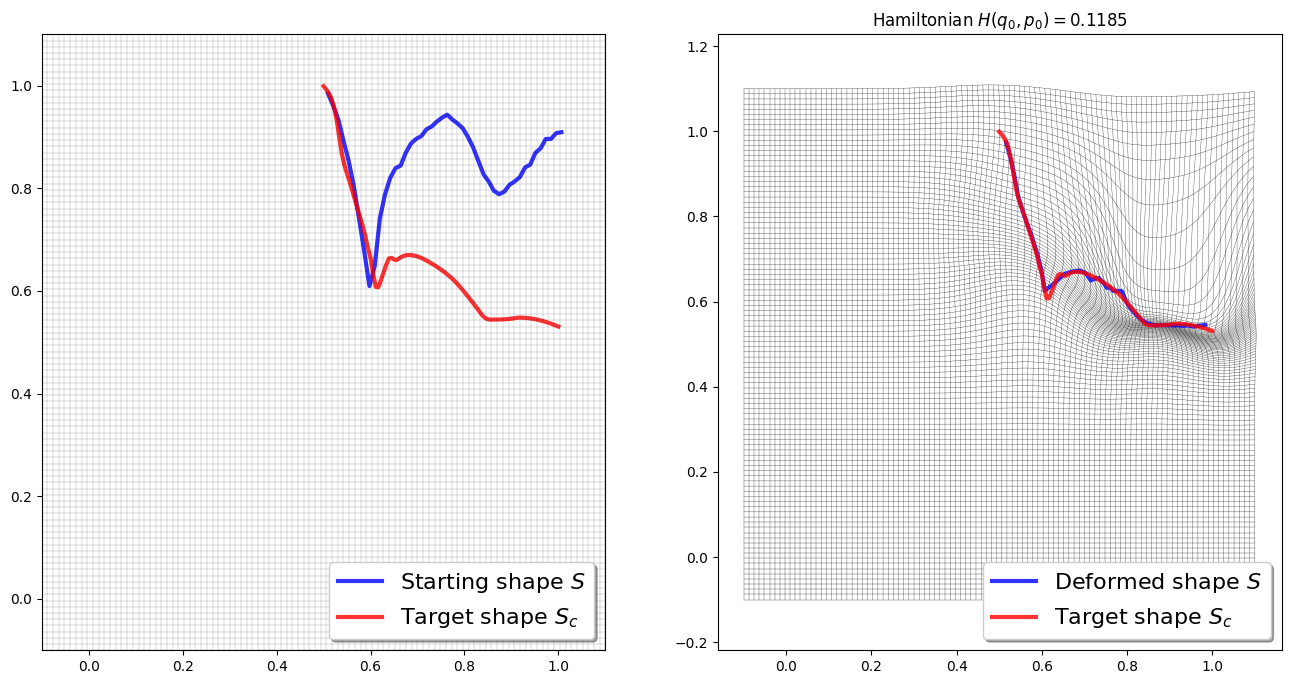}{Example of diffeomorphic registration for current-represented functional outputs.}{fig:plot_appariement_curves}

Starting from the experimental curves, we learn for each output $q^{(i)}$ the initial velocities $v_0^{(i)}$ in the geodesic shooting formulation. Then, we learn the relation between simulation parameters $\beta$ and $v_0$ with a Gaussian process regression with PCA. Other dimension reduction techniques could be performed, but PCA has the advantage of leaving the Gaussian process distribution tractable thanks to its linearity. We work with the initial velocity $v_0$ instead of the initial momentum $\pi_0$ in the surrogate model because the velocities are regularized by the kernel smoothing, which makes their inference easier.

Some metrics of performance for the PCA-GP model are given in Table \ref{tab:gp_pca_func}. The averaged 1D empirical coverage curve is displayed in Figure \ref{fig:empirical_coverage_func}.
\begin{table}[h]
\centering
\begin{tabular}{lc}
\hline
Validation metric \qquad &  \\
\hline
Aggregated NRMSE \qquad & \qquad $0.397$ \qquad \\
Aggregated NMAE  \qquad & \qquad $0.279$ \qquad \\
Aggregated $Q^2$ \qquad & \qquad $0.828$ \qquad \\
Aggregated CRPS  \qquad & \qquad $0.202$ \qquad \\
Aggregated IAE   \qquad & \qquad $0.060$ \qquad \\
RRMSE (\%)       \qquad & \qquad $42.92$ \qquad \\
\hline
\end{tabular}
\caption{Predictive performance of the GP-PCA surrogate model for Johnson’s damage calibration.}
\label{tab:gp_pca_func}
\end{table}

\fig{0.45}{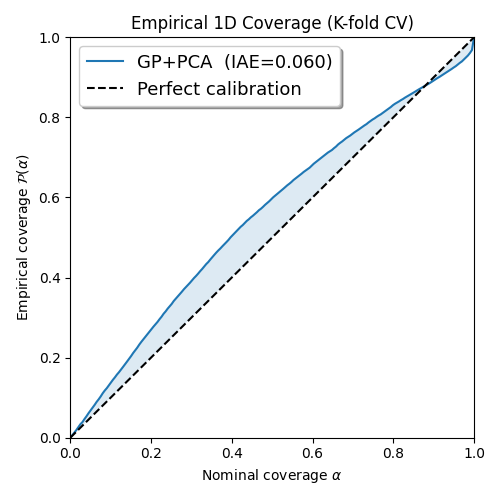}{Averaged empirical 1D coverage as a function of nominal coverage for the GP-PCA surrogate model.}{fig:empirical_coverage_func}

We first perform Bayesian calibration without a systematic model bias, as described in Section \ref{sec:prob_model}. In Figure \ref{fig:posterior_samples_curves}, we provide the maximum a posteriori and the credible interval of the predictive posterior distribution at confidence level $95 \%$.

\fig{0.7}{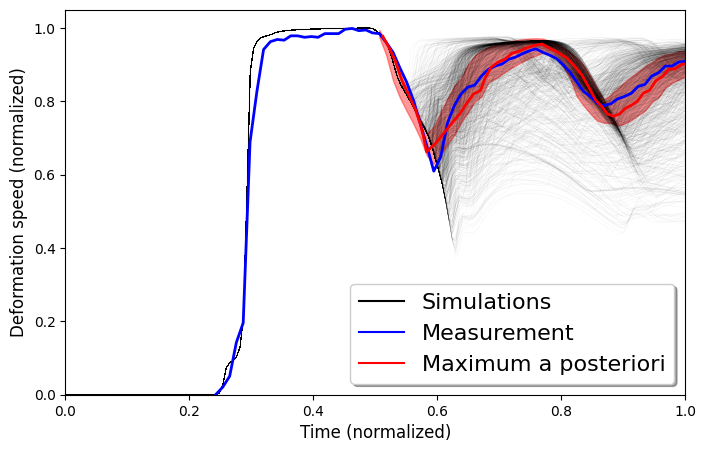}{Predictive posterior distribution and maximum a posteriori for the functional calibration of deformation speed with LDDMM distance}{fig:posterior_samples_curves}

Overall, the experimental curve lies within the predictive posterior credible intervals, but one can see non-negligible disparities around the first local minimum.

In Figure \ref{fig:predposterior_sm}, we provide a similar plot, only this time we account for the uncertainties of the GP-PCA surrogate model, as described in Section \ref{sec:prob_model_sm}. The predictive posterior is mostly unchanged, with slightly larger credible intervals. Thus, the surrogate model is not the main source of uncertainties here.

\fig{0.7}{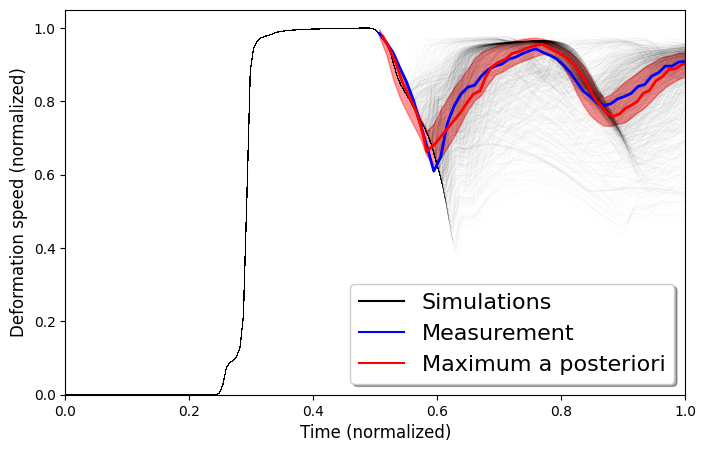}{Predictive posterior distribution and maximum a posteriori for the functional calibration of deformation speed with LDDMM distance, with inclusion of GP-PCA surrogate model error.}{fig:predposterior_sm}

Additionally, we observe similar disparities around the first local minimum. This leads us to believe the simulation has some systematic bias with regard to the physical phenomena involved. Consequently, we use the probabilistic model described in Section \ref{sec:prob_model_disc} to add a discrepancy. The predictive posterior is shown in Figure \ref{fig:predposterior_disc}.

\fig{0.7}{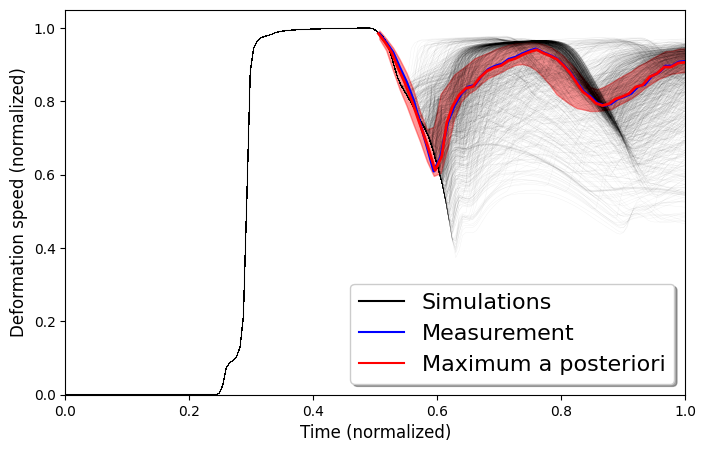}{Predictive posterior distribution and maximum a posteriori for the functional calibration of deformation speed with LDDMM distance, with discrepancy modeling.}{fig:predposterior_disc}

\begin{figure}[htbp]
\centering

\begin{subfigure}[t]{0.45\textwidth}
    \centering
    \includegraphics[width=\textwidth]{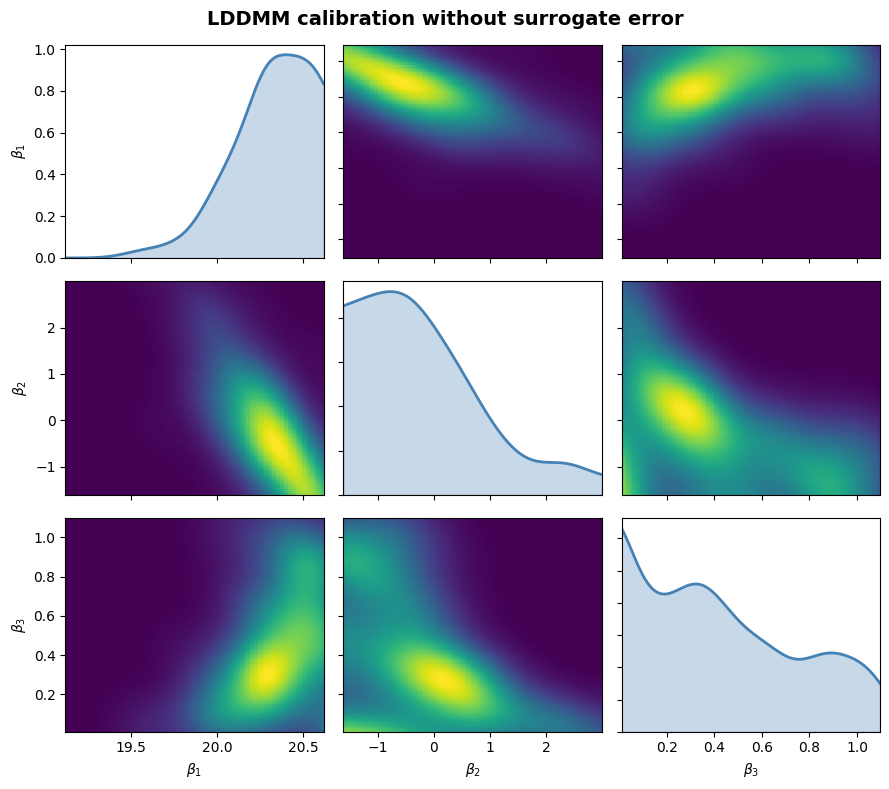}
    \caption{LDDMM calibration - without surrogate model error.}
    \label{fig:without_sm}
\end{subfigure}
\hfill
\begin{subfigure}[t]{0.45\textwidth}
    \centering
    \includegraphics[width=\textwidth]{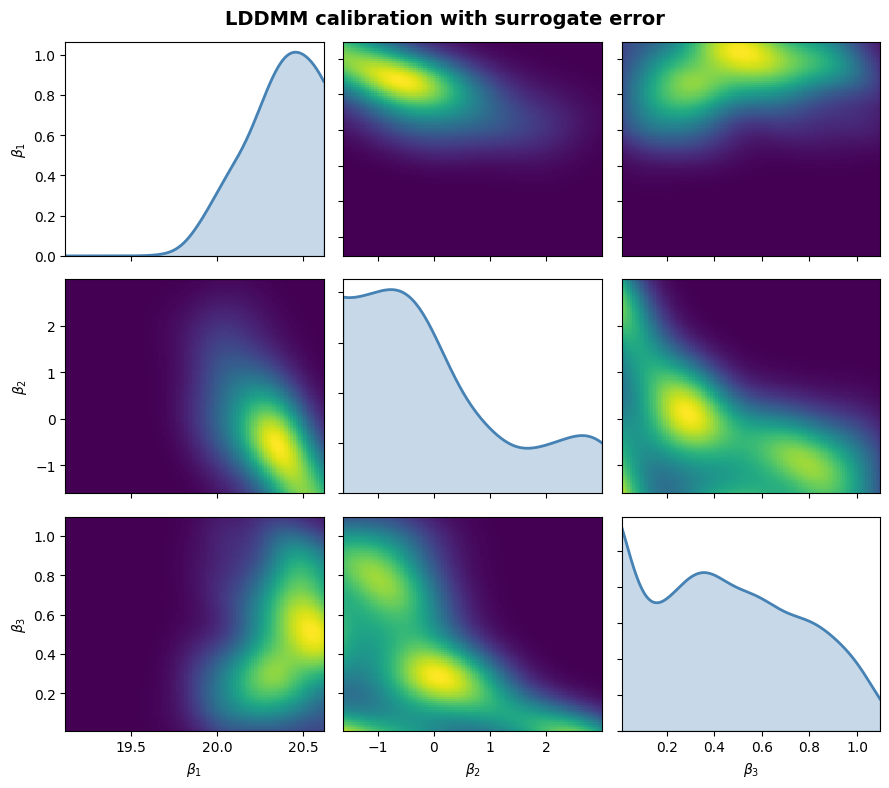}
    \caption{LDDMM calibration - with surrogate model error}
    \label{fig:with_sm}
\end{subfigure}

\vspace{0.5em}

\begin{subfigure}[t]{0.45\textwidth}
    \centering
    \includegraphics[width=\textwidth]{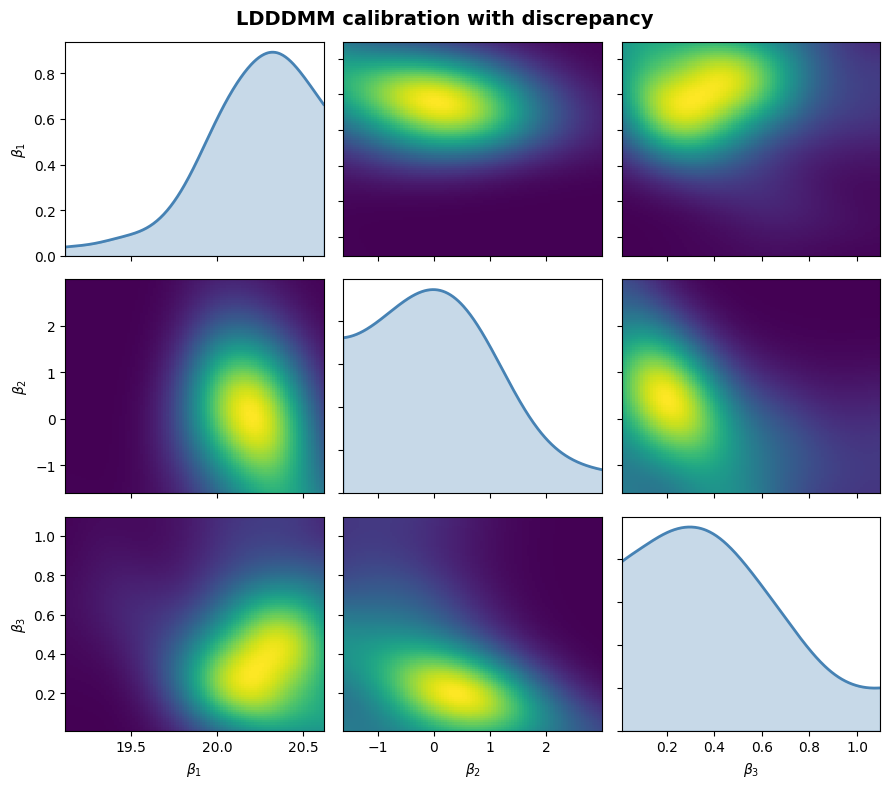}
    \caption{LDDMM calibration - with discrepancy and surrogate model error.}
    \label{fig:with_disc}
\end{subfigure}
\hfill
\begin{subfigure}[t]{0.45\textwidth}
    \centering
    \includegraphics[width=\textwidth]{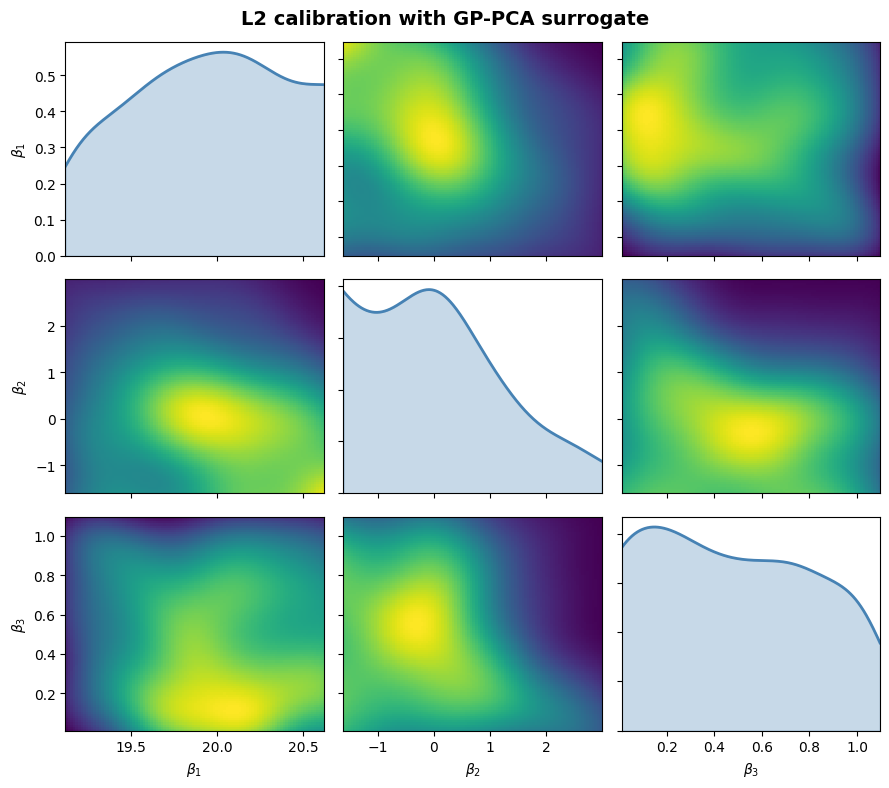}
    \caption{L2 calibration with GP-PCA surrogate.}
    \label{fig:with_L2}
\end{subfigure}

\caption{Comparison of all posterior densities $p(\beta | \qmes)$ for the different methods.}
\label{fig:posterior_comparison}

\end{figure}

The added discrepancy naturally resets the MAP onto the reference measurement. There is, however, some loss in identifiability between the effects of $\xi$ and $\beta$. To highlight this, we compare the posterior marginals on $\beta$ to the posterior obtained in the cases without discrepancy. The distributions are highlighted in Figure \ref{fig:posterior_comparison}.

These results first highlight the significant improvement over a standard L2 calibration approach which fails to identify the simulation parameters. Additionally, the effect of the GP-PCA surrogate error can be seen by the widening of the posterior distribution between Figure \ref{fig:without_sm} and \ref{fig:with_sm}. Still, the effect on the posterior distribution is small compared to the overall uncertainty.

We also observe the loss of identifiability introduced by the discrepancy term. While the posterior distribution in Figure \ref{fig:with_disc} is still much narrower than Figure \ref{fig:with_L2}, it is more spread-out than \ref{fig:without_sm} and \ref{fig:with_sm}. Part of the mismatch can be explained by the parameters $\xi$ in the discrepancy term, allowing for wider posterior marginals on $\beta$.

Overall, the proposed method offers much better results than a standard L2 loss with PCA dimension reduction and surrogate modeling. Moreover, it offers a unified framework able to deal with calibration tasks for objects such as images, probability measures, surfaces in $\mathbb{R}^3$, or functional graphs.

\newpage 
\section{Conclusion}
This work introduces large deformation distances in the Bayesian calibration framework. Our goal was to build a geometry-aware metric to compare high-dimensional outputs for calibration tasks. We showed that LDDMM theory can be adapted to Bayesian calibration tasks. The inference task boils down to inference in the Lie algebra, whose RKHS structure allows Bayesian modeling. It is also possible to add a discrepancy term to the probabilistic modeling in a similar fashion to a traditional Kennedy-O’Hagan model.

The Bayesian calibration of computer models with large-dimensional outputs is possible with this method, without requiring a large amount of data (as few as a hundred). It has been highlighted with a first toy example on $32 \times 32$ images, and on functional outputs with $n_{t_1} = 182$ time steps. It can be extended to higher resolution images. The main cost in computing time lies in the diffeomorphic registration of the simulations and the training of the GP-PCA surrogate model. These steps, however, are done once and occur before the MCMC sampling, making this method very tractable.

On top of this, for a low-data regime, we anticipate that most of the residual uncertainties would come from the uncertainty of the surrogate model, making it possible to use goal-oriented sequential design strategies to reduce the surrogate uncertainties and explore relevant simulation parameters with regard to the reference  $\qmes$ \cite{santner2003design, bect2019supermartingale}. These strategies have not yet been explored for diffeomorphic registration distances.

\newpage
\printbibliography

@article{kennedy2001bayesian,
  title     = {Bayesian calibration of computer models},
  author    = {Kennedy, Marc C and O'Hagan, Anthony},
  journal   = {Journal of the Royal Statistical Society: Series B (Statistical Methodology)},
  volume    = {63},
  number    = {3},
  pages     = {425--464},
  year      = {2001},
  publisher = {Wiley Online Library}
}

@article{miller2006geodesic,
  title={Geodesic shooting for computational anatomy},
  author={Miller, Michael I and Trouv{\'e}, Alain and Younes, Laurent},
  journal={Journal of mathematical imaging and vision},
  volume={24},
  number={2},
  pages={209--228},
  year={2006},
  publisher={Springer}
}

@article{miller2002metrics,
  title={On the metrics and Euler-Lagrange equations of computational anatomy},
  author={Miller, Michael I and Trouv{\'e}, Alain and Younes, Laurent},
  journal={Annual review of biomedical engineering},
  volume={4},
  number={1},
  pages={375--405},
  year={2002},
  publisher={Annual Reviews 4139 El Camino Way, PO Box 10139, Palo Alto, CA 94303-0139, USA}
}

@article{trouve2005local,
  title={Local geometry of deformable templates},
  author={Trouv{\'e}, Alain and Younes, Laurent},
  journal={SIAM journal on mathematical analysis},
  volume={37},
  number={1},
  pages={17--59},
  year={2005},
  publisher={SIAM}
}

@article{vialard2012diffeomorphic,
  title={Diffeomorphic 3D image registration via geodesic shooting using an efficient adjoint calculation},
  author={Vialard, Fran{\c{c}}ois-Xavier and Risser, Laurent and Rueckert, Daniel and Cotter, Colin J},
  journal={International Journal of Computer Vision},
  volume={97},
  number={2},
  pages={229--241},
  year={2012},
  publisher={Springer}
}

@article{charon2013varifold,
  title={The varifold representation of nonoriented shapes for diffeomorphic registration},
  author={Charon, Nicolas and Trouv{\'e}, Alain},
  journal={SIAM journal on Imaging Sciences},
  volume={6},
  number={4},
  pages={2547--2580},
  year={2013},
  publisher={SIAM}
}

@article{beg2005computing,
  title     = {Computing large deformation metric mappings via geodesic flows of diffeomorphisms},
  author    = {Beg, M Faisal and Miller, Michael I and Trouv{\'e}, Alain and Younes, Laurent},
  journal   = {International journal of computer vision},
  volume    = {61},
  number    = {2},
  pages     = {139--157},
  year      = {2005},
  publisher = {Springer}
}

@article{arguillere2015shape,
  title={Shape deformation analysis from the optimal control viewpoint},
  author={Arguillere, Sylvain and Tr{\'e}lat, Emmanuel and Trouv{\'e}, Alain and Younes, Laurent},
  journal={Journal de math{\'e}matiques pures et appliqu{\'e}es},
  volume={104},
  number={1},
  pages={139--178},
  year={2015},
  publisher={Elsevier}
}

@inproceedings{franccois2021metamorphic,
  title={Metamorphic image registration using a semi-Lagrangian scheme},
  author={Fran{\c{c}}ois, Anton and Gori, Pietro and Glaun{\`e}s, Joan},
  booktitle={International Conference on Geometric Science of Information},
  pages={781--788},
  year={2021},
  organization={Springer}
}

@article{holm2009euler,
  title={The Euler-Poincar{\'e} theory of metamorphosis},
  author={Holm, Darryl and Trouv{\'e}, Alain and Younes, Laurent},
  journal={Quarterly of Applied Mathematics},
  volume={67},
  number={4},
  pages={661--685},
  year={2009}
}

@article{trouve2005metamorphoses,
  title={Metamorphoses through lie group action},
  author={Trouv{\'e}, Alain and Younes, Laurent},
  journal={Foundations of computational mathematics},
  volume={5},
  number={2},
  pages={173--198},
  year={2005},
  publisher={Springer}
}

@article{mang2017lagrangian,
  title={A Lagrangian Gauss--Newton--Krylov solver for mass-and intensity-preserving diffeomorphic image registration},
  author={Mang, Andreas and Ruthotto, Lars},
  journal={SIAM Journal on Scientific Computing},
  volume={39},
  number={5},
  pages={B860--B885},
  year={2017},
  publisher={SIAM}
}

@article{trouve1998diffeomorphisms,
  title={Diffeomorphisms groups and pattern matching in image analysis},
  author={Trouv{\'e}, Alain},
  journal={International journal of computer vision},
  volume={28},
  number={3},
  pages={213--221},
  year={1998},
  publisher={Springer}
}

@article{chodosh2015discontinuity,
  title={On discontinuity of planar optimal transport maps},
  author={Chodosh, Otis and Jain, Vishesh and Lindsey, Michael and Panchev, Lyuboslav and Rubinstein, Yanir A},
  journal={Journal of Topology and Analysis},
  volume={7},
  number={02},
  pages={239--260},
  year={2015},
  publisher={World Scientific}
}

@article{higdon2008computer,
  title={Computer model calibration using high-dimensional output},
  author={Higdon, Dave and Gattiker, James and Williams, Brian and Rightley, Maria},
  journal={Journal of the American Statistical Association},
  volume={103},
  number={482},
  pages={570--583},
  year={2008},
  publisher={Taylor \& Francis}
}

@article{polette2025change,
  title={Change of measure for Bayesian field inversion with hierarchical hyperparameters sampling},
  author={Polette, Nad{\`e}ge and Le Ma{\^\i}tre, Olivier and Sochala, Pierre and Gesret, Alexandrine},
  journal={Journal of Computational Physics},
  volume={529},
  pages={113888},
  year={2025},
  publisher={Elsevier}
}

@article{conti2010bayesian,
  title={Bayesian emulation of complex multi-output and dynamic computer models},
  author={Conti, Stefano and O’Hagan, Anthony},
  journal={Journal of statistical planning and inference},
  volume={140},
  number={3},
  pages={640--651},
  year={2010},
  publisher={Elsevier}
}

@article{francom2025elastic,
  title={Elastic Bayesian model calibration},
  author={Francom, Devin and Tucker, J Derek and Huerta, Gabriel and Shuler, Kurtis and Ries, Daniel},
  journal={SIAM/ASA Journal on Uncertainty Quantification},
  volume={13},
  number={1},
  pages={195--227},
  year={2025},
  publisher={SIAM}
}

@article{qian2014,
author = {Qian Xie and Sebastian Kurtek and Anuj Srivastava},
title = {{Analysis of AneuRisk65 data: Elastic shape registration of curves}},
volume = {8},
journal = {Electronic Journal of Statistics},
number = {2},
publisher = {Institute of Mathematical Statistics and Bernoulli Society},
pages = {1920 -- 1929},
year = {2014},
doi = {10.1214/14-EJS938D},
URL = {https://doi.org/10.1214/14-EJS938D}
}

@article{cheng2016bayesian,
author = {Cheng, Wen and Dryden, Ian and Huang, Xianzheng},
year = {2016},
month = {06},
pages = {},
title = {Bayesian Registration of Functions and Curves},
volume = {11},
journal = {Bayesian Analysis},
doi = {10.1214/15-BA957}
}

@article{ceritoglu2009multi,
  title={Multi-contrast large deformation diffeomorphic metric mapping for diffusion tensor imaging},
  author={Ceritoglu, Can and Oishi, Kenichi and Li, Xin and Chou, Ming-Chung and Younes, Laurent and Albert, Marilyn and Lyketsos, Constantine and van Zijl, Peter CM and Miller, Michael I and Mori, Susumu},
  journal={Neuroimage},
  volume={47},
  number={2},
  pages={618--627},
  year={2009},
  publisher={Elsevier}
}

@article{charlier2017fshape,
  title={The fshape framework for the variability analysis of functional shapes},
  author={Charlier, Benjamin and Charon, Nicolas and Trouv{\'e}, Alain},
  journal={Foundations of Computational Mathematics},
  volume={17},
  number={2},
  pages={287--357},
  year={2017},
  publisher={Springer}
}

@article{charon2014functional,
  title={Functional currents: a new mathematical tool to model and analyse functional shapes},
  author={Charon, Nicolas and Trouv{\'e}, Alain},
  journal={Journal of mathematical imaging and vision},
  volume={48},
  number={3},
  pages={413--431},
  year={2014},
  publisher={Springer}
}

@article{marrel2024probabilistic,
  title={Probabilistic surrogate modeling by Gaussian process: A new estimation algorithm for more robust prediction},
  author={Marrel, Amandine and Iooss, Bertrand},
  journal={Reliability Engineering \& System Safety},
  volume={247},
  pages={110120},
  year={2024},
  publisher={Elsevier}
}

@article{stuart2010inverse,
  title     = {Inverse problems: a Bayesian perspective},
  author    = {Stuart, Andrew M},
  journal   = {Acta numerica},
  volume    = {19},
  pages     = {451--559},
  year      = {2010},
  publisher = {Cambridge University Press}
}

@inproceedings{durrleman2011optimal,
  title        = {Optimal data-driven sparse parameterization of diffeomorphisms for population analysis},
  author       = {Durrleman, Sandy and Prastawa, Marcel and Gerig, Guido and Joshi, Sarang},
  booktitle    = {Biennial International Conference on Information Processing in Medical Imaging},
  pages        = {123--134},
  year         = {2011},
  organization = {Springer}
}

@inproceedings{perrin2018calibration,
  title={Calibration of Johnson’s damage model by a Bayesian approach},
  author={Perrin, Guillaume and Pillon, Laurianne},
  booktitle={EPJ Web of Conferences},
  volume={183},
  pages={01037},
  year={2018},
  organization={EDP Sciences}
}

@article{hoffman2014no,
  title={The No-U-Turn sampler: adaptively setting path lengths in Hamiltonian Monte Carlo.},
  author={Hoffman, Matthew D and Gelman, Andrew and others},
  journal={J. Mach. Learn. Res.},
  volume={15},
  number={1},
  pages={1593--1623},
  year={2014}
}

@article{girolami2011riemann,
  title={Riemann manifold langevin and hamiltonian monte carlo methods},
  author={Girolami, Mark and Calderhead, Ben},
  journal={Journal of the Royal Statistical Society Series B: Statistical Methodology},
  volume={73},
  number={2},
  pages={123--214},
  year={2011},
  publisher={Oxford University Press}
}

@article{kingma2014adam,
  title={Adam: A method for stochastic optimization},
  author={Kingma, Diederik P and Ba, Jimmy},
  journal={arXiv preprint arXiv:1412.6980},
  year={2014}
}

@article{bingham2018pyro,
author = {Bingham, Eli and Chen, Jonathan P. and Jankowiak, Martin and Obermeyer, Fritz and
          Pradhan, Neeraj and Karaletsos, Theofanis and Singh, Rohit and Szerlip, Paul and
          Horsfall, Paul and Goodman, Noah D.},
title = {{Pyro: Deep Universal Probabilistic Programming}},
journal = {Journal of Machine Learning Research},
year = {2018}
}

@book{younes2010shapes,
  title={Shapes and diffeomorphisms},
  author={Younes, Laurent},
  volume={171},
  year={2010},
  publisher={Springer}
}

@book{zeidler2013nonlinear,
  title={Nonlinear functional analysis and its applications: III: variational methods and optimization},
  author={Zeidler, Eberhard},
  year={2013},
  publisher={Springer Science \& Business Media}
}

@article{duane1987hybrid,
  title={Hybrid monte carlo},
  author={Duane, Simon and Kennedy, Anthony D and Pendleton, Brian J and Roweth, Duncan},
  journal={Physics letters B},
  volume={195},
  number={2},
  pages={216--222},
  year={1987},
  publisher={Elsevier}
}

@article{neal1992bayesian,
  title={Bayesian learning via stochastic dynamics},
  author={Neal, Radford},
  journal={Advances in neural information processing systems},
  volume={5},
  year={1992}
}

@article{storlie2015calibration,
  title={Calibration of computational models with categorical parameters and correlated outputs via Bayesian smoothing spline ANOVA},
  author={Storlie, Curtis B and Lane, William A and Ryan, Emily M and Gattiker, James R and Higdon, David M},
  journal={Journal of the American Statistical Association},
  volume={110},
  number={509},
  pages={68--82},
  year={2015},
  publisher={Taylor \& Francis}
}

@book{congdon2005bayesian,
  title={Bayesian models for categorical data},
  author={Congdon, Peter},
  year={2005},
  publisher={John Wiley \& Sons}
}

@book{santner2003design,
  title={The design and analysis of computer experiments},
  author={Santner, Thomas J and Williams, Brian J and Notz, William I and Williams, Brain J},
  volume={1},
  year={2003},
  publisher={Springer}
}

@article{bect2019supermartingale,
  title={A supermartingale approach to Gaussian process based sequential design of experiments},
  author={Bect, Julien and Bachoc, Fran{\c{c}}ois and Ginsbourger, David},
  journal={Bernoulli},
  volume={25},
  number={4A},
  pages={2883--2919},
  year={2019},
  publisher={JSTOR}
}

@inproceedings{cherief2020mmd,
  title={MMD-Bayes: Robust Bayesian estimation via maximum mean discrepancy},
  author={Ch{\'e}rief-Abdellatif, Badr-Eddine and Alquier, Pierre},
  booktitle={Symposium on Advances in Approximate Bayesian Inference},
  pages={1--21},
  year={2020},
  organization={PMLR}
}

@inproceedings{gross1967abstract,
  title={Abstract Wiener spaces},
  author={Gross, Leonard},
  booktitle={Proceedings of the Fifth Berkeley Symposium on Mathematical Statistics and Probability, Volume 2: Contributions to Probability Theory, Part 1},
  volume={5.2A},
  year={1967},
  pages={31--43}
}

@book{kuo2006gaussian,
  title={Gaussian measures in Banach spaces},
  author={Kuo, Hui-Hsiung},
  year={1975},
  publisher={Springer}
}

@article{kadri2016operator,
  title={Operator-valued kernels for learning from functional response data},
  author={Kadri, Hachem and Duflos, Emmanuel and Preux, Philippe and Canu, St{\'e}phane and Rakotomamonjy, Alain and Audiffren, Julien},
  journal={Journal of Machine Learning research},
  volume={17},
  number={20},
  pages={1--54},
  year={2016}
}

\end{document}